\documentclass[conference]{IEEEtran}

\usepackage{cite}
\usepackage{amsmath,amssymb,amsfonts}
\usepackage{algorithmic}
\usepackage{graphicx}
\usepackage{textcomp}
\usepackage{xcolor}
\def\BibTeX{{\rm B\kern-.05em{\sc i\kern-.025em b}\kern-.08em
    T\kern-.1667em\lower.7ex\hbox{E}\kern-.125emX}}

\usepackage{url}
\usepackage{graphicx}
\usepackage{amsmath}
\usepackage{amsthm}
\usepackage{dsfont}
\usepackage{subcaption}
\usepackage{color}
\usepackage{enumerate}
\usepackage{cases}
\usepackage{multicol}
\usepackage{multirow}
\usepackage{enumitem}
\usepackage{bm}
\usepackage{import}
\usepackage{lscape}
\usepackage{amssymb}
\usepackage{xcolor}
\usepackage{float}
\usepackage{booktabs}
\usepackage{hyperref}

\usepackage[algo2e,linesnumbered,ruled,vlined]{algorithm2e}
\usepackage{caption}
\usepackage{wrapfig}

\definecolor{BrightBlue}{RGB}{65, 145, 225}

\definecolor{figure_blue}{RGB}{53, 132, 187}
\definecolor{figure_orange}{RGB}{255, 139, 38}
\definecolor{figure_green}{RGB}{65, 169, 65}
\definecolor{figure_red}{RGB}{218, 60, 61}
\definecolor{figure_purple}{RGB}{158, 118, 195}
\definecolor{figure_brown}{RGB}{160, 82, 45}
\definecolor{figure_pink}{RGB}{255,105,180}

\makeatletter
\newcommand\footnoteref[1]{\protected@xdef\@thefnmark{\ref{#1}}\@footnotemark}
\makeatother

\DeclareMathOperator{\sign}{sign}

\begin{document}

\title{LIA: Privacy-Preserving Data Quality Evaluation in Federated Learning Using a Lazy Influence Approximation}

\author{
\IEEEauthorblockN{Ljubomir Rokvic\IEEEauthorrefmark{1},
Panayiotis Danassis\IEEEauthorrefmark{2},
Sai Praneeth Karimireddy\IEEEauthorrefmark{3} and
Boi Faltings\IEEEauthorrefmark{1}}
\IEEEauthorblockA{\IEEEauthorrefmark{1}\'Ecole Polytechnique F\'ed\'erale de Lausanne (EPFL), Switzerland}
\IEEEauthorblockA{\IEEEauthorrefmark{2}Telenor Research, Norway}
\IEEEauthorblockA{\IEEEauthorrefmark{3}University of Southern California, USA}
\IEEEauthorblockA{ljubomir.rokvic@epfl.ch, panayiotis.danassis@telenor.com, karimire@usc.edu, boi.faltings@epfl.ch}
}

\maketitle

\begin{abstract}
In Federated Learning, it is crucial to handle low-quality, corrupted, or malicious data. However, traditional data valuation methods are not suitable due to privacy concerns. To address this, we propose a simple yet effective approach that utilizes a new influence approximation called \emph{"lazy influence"} to filter and score data while preserving privacy. To do this, each participant uses their own data to estimate the influence of another participant's batch and sends a differentially private obfuscated score to the central coordinator. Our method has been shown to \emph{successfully filter out biased and corrupted data} in various simulated and \emph{real-world} settings, achieving a recall rate of over $>90\%$ (sometimes up to $100\%$) while maintaining \emph{strong differential privacy} guarantees with $\varepsilon \leq 1$.
\end{abstract}

\begin{IEEEkeywords}
Federated Learning, Data Filtering, Data Poisoning, Robustness, Privacy.
\end{IEEEkeywords}

\section{Introduction} \label{sec: Introduction}

The success of Machine Learning (ML) depends to a large extent on the availability of high-quality data~\cite{oala2023dmlr}. Having the ability to \emph{score} and \emph{filter} irrelevant, noisy, or malicious data can significantly improve model accuracy and speed up training. However, ensuring data quality is especially challenging in Federated Learning (FL)~\cite{mcmahan2017communication,kairouz2021advances,wang2021field}. In FL, a single \emph{Center} uses data from independent and sometimes self-interested \emph{data holders} to train a model jointly. Data holders often operate resource-constrained edge devices and include businesses and medical institutions that must protect the privacy of their data due to confidentiality or legal constraints. Thus, we never have direct access to the training data in order to inspect it, and privacy is a strong concern. This raises our main question: \emph{how can we evaluate data quality in a distributed manner while preserving privacy?}

Imagine a collaborative project involving multiple hospitals across the globe, each with its own repository of chest X-ray scans of patients. The goal is to use FL to train a robust ML model capable of diagnosing heart conditions from these X-rays~\cite{rajpurkar2017chexnet}. However, each hospital adheres to different data collection protocols. E.g., Hospital A may use a different scanner making it incompatible with the rest of the data. If included, it can lead to a large drop in accuracy~\cite{liu2019comparison}. More insidiously, Hospital B may habitually include textual annotations directly on the scans for internal communication. The model may rely on these shortcuts instead of looking at the actual scan~\cite{oakden2020hidden}. 
At the same time, it is important not to discard data about rare diseases, even if it might appear very different from most other data.

To address these issues, we propose a ``rite of passage'' mechanism where a hospital's data is admitted into the FL system only if no issues are detected. By this, we not only ensure the integrity of the FL procedure, but can also flag potentially bad data-collection procedures for the hospitals. However, note that patient X-rays are very sensitive information and protected by strong privacy regulations such as HIPAA. Thus we need to ensure that both the model training and (for our purpose) more importantly the data quality assessment is performed in a federated and privacy preserving manner minimizing the patient information being leaked.

To prevent such a scheme from filtering out valuable data about rare conditions, quality assessment cannot be based on simple data similarity, which would consider them outliers.
A clean way of quantifying the quality of data point(s) is via the notion of \emph{influence}~\cite{pmlrv70koh17a,cook1980characterizations}. Intuitively, influence quantifies the marginal contribution of a data point (or batch of points) on a model's accuracy. One can compute this by comparing the difference in the model's empirical risk when trained with and without the point in question. 
While the influence metric can be highly informative, it is impractical to compute: re-training a model is time-consuming, costly, and often impossible, as participants do not have access to the entire dataset. We propose a simple and practical approximation of the \emph{sign} of the exact influence (\emph{Lazy Influence Approximation}) which is based on an estimate of the direction of the model after a small number of local training epochs with the new data.

\begin{figure*}
    \centering
    \includegraphics[width=0.9\linewidth, clip, trim={0em 2.975cm 0em 0em}]{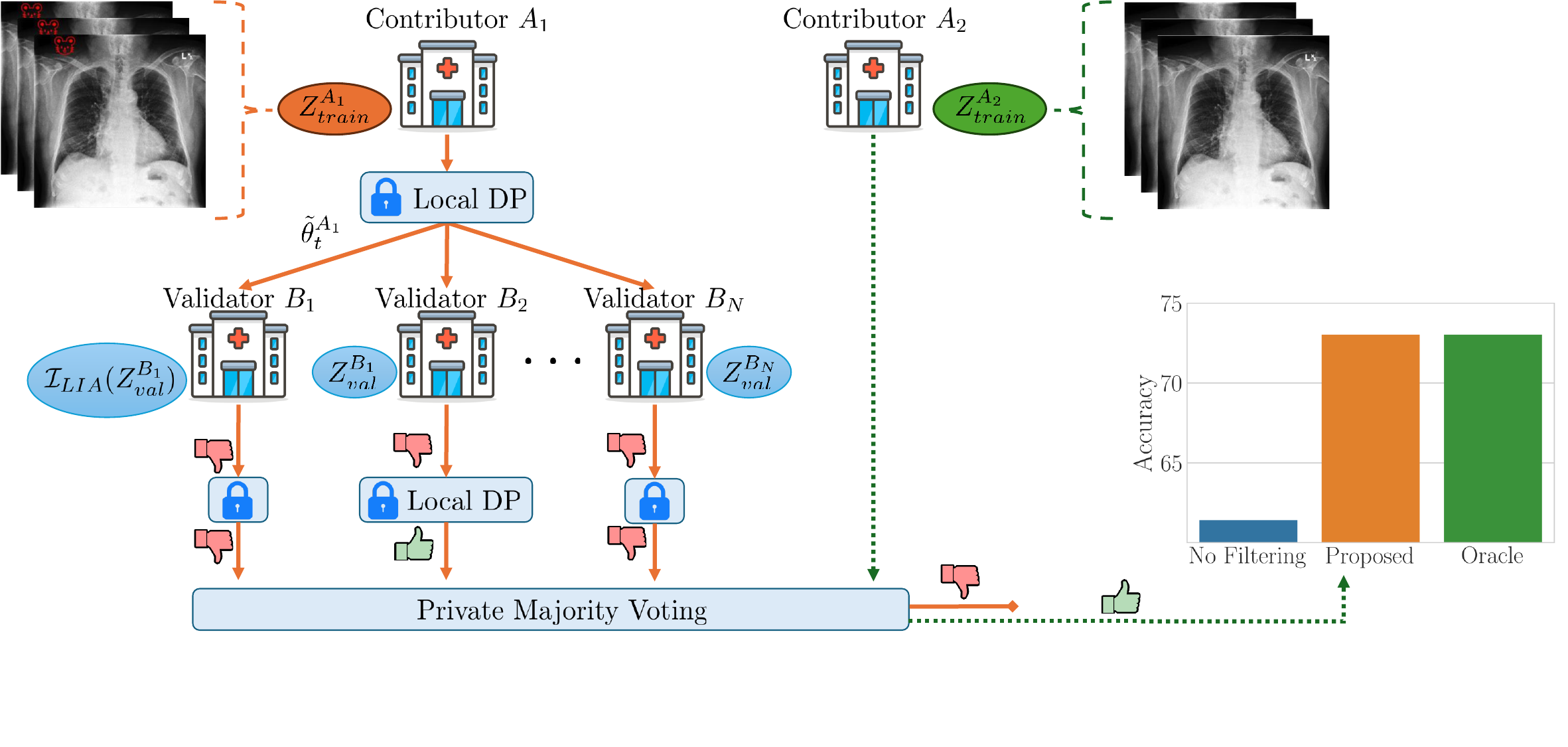}
    \caption{Data filtering procedure. Data contributors $A_1$ and $A_2$ want to join the federation, but might have biased or corrupted data (e.g., watermarks on $A_1$ X-rays). $A_1$ sends a differentially private (DP) partially updated model ($\tilde{\theta}_t^{A_1}$) to validators $B_i$, who submit a DP vote based on the performance they observe when using the model $\tilde{\theta}_t^{A_1}$ on their task (\emph{Lazy Influence Approximation}). The aggregated votes are used as a \emph{`rite of passage'}, to decide whether to incorporate $A_1$'s data. The same process happens for $A_2$. $A_2$ is accepted in the federation, while $A_1$ is filtered out. Filtering can significantly improve the model's accuracy from 61\% (no filtering) to 73\% (proposed), matching the performance of the optimal (oracle) filtering on diagnosing heart conditions from X-rays using \emph{real-data} from~\cite{rajpurkar2017chexnet}.}
    \label{figure:1}
\end{figure*}

The second challenge is to limit the information leaked by this approximate influence measure  to preserve the privacy of the data.
Many approaches to Federated Learning 
(e.g.,~\cite{brendan2018learning,DBLP:conf/bigdataconf/TriastcynF19}) remedy this by combining FL with Differential Privacy (DP)~\cite{dwork2006,dwork2006our,dwork2006calibrating,kairouz2021practical,de2204unlocking,choquette2022multi}, a 
data anonymization technique that many researchers view as the gold standard~\cite{triastcyn2020data}. 
We show how the sign of influence can be approximated in an FL setting while maintaining strong differential privacy guarantees. Specifically, there are two sets of participants' data that we need to protect: the training and the validation data (see also Section~\ref{sec: High-Level Description of Our Setting}). We use DP-SGD for this training to ensure datapoint-level local DP~\cite{abadi2016deep,opacus}
To ensure the privacy of the validation data and the influence approximation itself, we employ a differentially private defense mechanism based on the idea of randomized response~\cite{warner1965randomized} (inspired by~\cite{erlingsson2014rappor}).  Together the two mechanisms ensure strong, \emph{worst-case differential privacy} guarantees while allowing for accurate data filtering. See Figure~\ref{figure:1} for an overview of our approach.

\subsection{Our Contributions} \label{sec: Our Contributions}

We address two major challenges in this work: (i) efficiently estimating the quality of a batch of training data, and (ii) keeping both the training and validation data used for this estimate private. For the former, we develop a novel metric called \emph{Lazy Influence Approximation (LIA)}, while for the latter, we combine DP-SGD and a differentially-private voting scheme. More specifically:

\textbf{(1)} We present a novel technique (\textbf{Lazy Influence Approximation}) for scoring and filtering data in Federated Learning.

\textbf{(2)} We show that our proposed distributed influence aggregation scheme allows for \textbf{robust scoring, even under rigorous, worst-case differential privacy guarantees} (privacy cost $\varepsilon < 1$). This is the recommended value in DP literature and much smaller than many other AI or ML applications.\footnote{AI or ML applications often assume $\varepsilon$ as large as 10~\cite{DBLP:conf/bigdataconf/TriastcynF19} (see, e.g.,~\cite{tang2017privacy}). For specific attacks, $\varepsilon = 10$ means that an adversary can theoretically reach an accuracy of $99.99\%$~\cite{DBLP:conf/bigdataconf/TriastcynF19}}.

\textbf{(3)} We evaluate our approach on four well-established datasets: CIFAR10, CIFAR100, \emph{real-data} on Human Activity Recognition on edge devices~\cite{anguita2013public}, and \emph{real-data} on diagnosing heart conditions from X-rays~\cite{rajpurkar2017chexnet}. Our simulations include \textbf{two different modalities} (image and sensor data), and \textbf{corruption of both the input $X$, and the labels $y$}. We demonstrate that \textbf{filtering using our scheme can eliminate the adverse effects of inaccurate data}. We also show it can be easily combined with other robust-FL training methods, to further enhance the model's performance. 

\subsection{High-Level Description of Our Setting} \label{sec: High-Level Description of Our Setting}

A center $C$ coordinates a set of participants that contribute their data to train a single model (Figure~\ref{figure:1}). $C$ has a small set of `warm-up' data, which are used to train an initial model $M_0$ that captures the desired input/output relation. 
The model is then updated with the contributions of the participants. We assume that each data holder participant has a set of training points that will be used to improve the model and a set of validation points that will be used to evaluate other participants' contributions. It must be kept private to prohibit participants from tailoring their contributions to the validation data. Each data holder participant can assume two roles: the role of the contributor ($A$) and the role of the validator ($B$). As a contributor, a participant performs a small number of local epochs to $M_t$ -- enough to get an estimate of the gradient\footnote{The number of local epochs is a hyperparameter. We do not need to train the model fully (e.g., 2-5 epochs).} -- using a batch of his training data $z_{A, t}$. Subsequently, $A$ sends the updated partial model $M_{t, A}$, with specifically crafted noise to ensure DP, to the validators. The applied noise protects the updated gradient while still retaining information on the usefulness of data. Each validator $B$ uses its validation dataset to approximate the empirical risk of $A$'s training batch (i.e., the approximate influence). This is done by evaluating each validation point and comparing the loss. In an FL setting, we can not re-train the model to compute the exact influence; instead, $B$ performs only a small number of training epochs, enough to estimate the direction of the model (Lazy Influence Approximation). 
As such, we look at the sign of the approximate influence (and not the magnitude). Each validator aggregates the signs of the influence for each validation point, applies controlled noise to ensure DP, and sends this information to the center. Finally, the center accepts $A$'s training batch if most of the $B$s report positive influence and rejects otherwise (majority voting). To recognize the value of data about rare conditions, it is important to use a broad sample of validation data. Following the federated learning idea, we distribute evaluation among many participants. Anyone capable of using the model can be a validator, so there could be many more than contributors.

\section{Related Work and Discussion} \label{sec: Related Work}

\paragraph{Federated Learning}

Federated Learning (FL)~\cite{mcmahan2017communication,kairouz2021advances,wang2021field,li2020federated} has emerged as an alternative method to train ML models on data obtained by many different agents. In FL, a center coordinates agents who acquire data and provide model updates. FL has been receiving increasing attention in both academia~\cite{lim2020federated,yang2019federated,he2020fedml,caldas2018leaf} and industry~\cite{hard2018federated,chen2019federated}, with a plethora of real-world applications.

\paragraph{Influence functions}

Influence functions are a standard method from robust statistics~\cite{cook1980characterizations} (see also Section \ref{sec: Methodology}), which were recently used as a method of explaining the predictions of black-box models \cite{pmlrv70koh17a}. They have also been used in the context of fast cross-validation in kernel methods and model robustness \cite{liu2014efficient,christmann2004robustness}. While a powerful tool, computing the influence involves too much computation and communication, and it requires access to the training and validation data (see~\cite{pmlrv70koh17a} and Section \ref{sec: Methodology}). There has also been recent work trying to combine Federated Learning with influence functions~\cite{xue2021toward}, though to the best of our knowledge, we are the first to provide a privacy-preserving alternative.

\paragraph{Data Filtering}

A common but computationally expensive approach for filtering in ML is to use the Shapley Value of the Influence to evaluate the quality of data~\cite{jia2019towards,ghorbani2019data,jia2019efficient,yan2020evaluating,pmlr-v97-ghorbani19c,watson2022differentially}. Other work includes, for example, rule-based filtering of least influential points~\cite{ogawa2013safe}, or constructing weighted data subsets (corsets)~\cite{dasgupta2009sampling}. Because of the privacy requirements in FL, contributed data is not directly accessible for assessing its quality.  \cite{tuor2021overcoming} propose a decentralized filtering process specific to federated learning, yet they do not provide any formal privacy guarantees. While data filtering might not always pose a significant problem in traditional ML, in an FL setting, it is more important because even a small percentage of mislabeled data can result in a significant drop in the combined model's accuracy.

\paragraph{Client Selection and Attack Detection}

Our setting can also be interpreted as potentially adversarial, but it should not be confused with Byzantine robustness. We do not consider threat scenarios as described in~\cite{cao2020fltrust} and~\cite{so2020byzantine}, where participants carefully craft malicious updates. Instead, we assume that the data used for those updates might be corrupt. For completeness and in lack of more relevant baselines, we compare our work to two Byzantine robust methods: KRUM~\cite{KRUM}, Trimmed-mean~\cite{pmlr-v80-yin18a}, and Centered-Clipping~\cite{karimireddy2020learning} (along to an oracle filter). These methods, though, require gradients to be transmitted as is, i.e., they lack any formal privacy guarantee to the participants' training data.
Furthermore, both of these techniques require the center to know the number of malicious participants a priori. Nevertheless, we show that our approach can be combined with such robust-FL training methods, to further enhance the model’s performance.

\paragraph{Differential Privacy}

Differential Privacy (DP)~\cite{dwork2006,dwork2006our,dwork2006calibrating} has emerged as the de facto standard for protecting the privacy of individuals. Informally, DP captures the increased risk to an individual's privacy incurred by participating in the learning process. Consider a participant being surveyed on a sensitive topic as a simplified, intuitive example. To achieve differential privacy, one needs a source of randomness; thus, the participant decides to flip a coin. Depending on the result (heads or tails), the participant can reply truthfully or randomly. An attacker can not know if the decision was taken based on the participant's preference or due to the coin toss. Of course, to get meaningful results, we need to bias the coin toward the actual data. In this simple example, the logarithm of the ratio $Pr[\text{heads}] / Pr[\text{tails}]$ represents the privacy cost (also referred to as the privacy budget), denoted traditionally by $\varepsilon$. 
Yet, one must be careful in designing a DP mechanism, as it is often hard to practically achieve a meaningful privacy guarantee (i.e., avoid adding a lot of noise and maintain high accuracy)~\cite{DBLP:conf/bigdataconf/TriastcynF19,DanassisPalma}. A variation of DP, instrumental in our context, given the decentralized nature of federated learning, is Local Differential Privacy (LDP)~\cite{kasiviswanathan2011can,dwork2014algorithmic}. LDP is a generalization of DP that assumes all communication between the data providers and the Center is public. Hence, we do not rely on the Center to take appropriate privacy measures and instead try limit information leakage in all communication. This allows us to work with untrusted central coordinators. An orthogonal concept is datapoint-level vs. user-level DP~\cite{wang2019beyond,augenstein2019generative,liu2020learning}. In user-level privacy, we attempt to protect datasets at the level of the data provider, rather than at the level of data-point. While this is typical in the cross-device FL where each data-provider represents the data of a single user, this does not make sense for multi-institutional collaboration~\cite{liu2022privacy}. In the setting we consider, each hospital (data provider) aggregates data from many patients. Thus, each user we want to protect is associated with a single data point, making datapoint-level DP more relevant. Thus, we consider datapoint-level local DP.
We assume that the participants and the Center are {\em honest but curious}, i.e., they don't actively attempt to corrupt the protocol but will try to learn about each other's data.

\section{Methodology}  \label{sec: Methodology}

We aim to address two challenges: (i) approximating the influence of a (batch of) data point(s) without having to re-train the entire model from scratch and (ii) doing so while protecting the privacy of the training and validation data. 
The latter is essential not only to protect users' sensitive information but also to ensure that malicious participants can not tailor their contributions to the validation data. 
We first introduce the notion of \emph{influence}~\cite{cook1980characterizations} and our proposed lazy approximation. Second, we describe a differentially private voting scheme for crowdsourcing the approximate influence values.

\paragraph{Setting}  \label{sec: Setting}

We consider a classification problem from some input space $\mathcal{X}$ (e.g., features, images, etc.) to an output space $\mathcal{Y}$ (e.g., labels). In a FL setting, there is a center $C$ that wants to learn a model $M(\theta)$ parameterized by $\theta \in \Theta$, with a non-negative loss function $L(z, \theta)$ on a sample $z=(\bar{x},y) \in \mathcal{X} \times \mathcal{Y}$. Let $R(Z, \theta) = \frac{1}{n}\sum_{i=1}^n L(z_i, \theta)$ denote the empirical risk, given a set of data $Z = \{z_i\}_{i=1}^n$. We assume that the empirical risk is differentiable in $\theta$.

\paragraph{Definitions}

In simple terms, influence measures the marginal contribution of a data point on a model's accuracy. A positive influence value indicates that a data point improves model accuracy, and vice-versa. More specifically, let $Z = \{z_i\}_{i=1}^n$,  $Z_{+j} = Z \cup z_j$ where $z_j \not\in Z$, and let $\hat{R} = \min_{\theta} R(Z, \theta)$ and $\hat{R}_{+j} = \min_{\theta} R(Z_{+j}, \theta)$, where $\hat{R}$ and $\hat{R}_{+j}$ denote the minimum empirical risk of their respective set of data. The \emph{influence} of datapoint $z_j$ on $Z$ is defined as $\mathcal{I}(z_j,Z) \triangleq \hat{R} - \hat{R}_{+j}$.

\subsection{Shortcomings of the Exact and Approximate Influence in a FL Setting}  \label{sec: Influence}

Despite being highly informative, influence functions have not achieved widespread use in FL (or ML in general). This is mainly due to the computational cost. The exact influence requires complete retraining of the model, which is time-consuming and very costly, especially for state-of-the-art, large ML models (importantly for our setting, we do not have direct access to the training data). Recently, the first-order Taylor approximation of influence~\cite{pmlrv70koh17a} (based on~\cite{cook1982residuals}) has been proposed as a practical method to understanding the effects of training points on the predictions of a \emph{centralized} ML model. While it can be computed without having to re-train the model, according to the following equation $\mathcal{I}_{appr}(z_{j}, z_{val}) \triangleq - \nabla_\theta L(z_{val}, \hat{\theta}) H^{-1}_{\hat{\theta}} \nabla_\theta L(z_j, \hat{\theta})$, it is still ill-matched for FL models for several key reasons.

Computing the influence approximation of~\cite{pmlrv70koh17a} requires \emph{forming and inverting} the Hessian of the empirical risk. With $n$ training points and $\theta \in \mathbb{R}^m$, this requires $O(nm^2 + m^3)$ operations~\cite{pmlrv70koh17a}, which is \emph{impractical} for modern-day deep neural networks with millions of parameters. To overcome these challenges,~\cite{pmlrv70koh17a} used implicit Hessian-vector products (HVPs) to more efficiently approximate $\nabla_\theta L(z_{val}, \hat{\theta}) H^{-1}_{\hat{\theta}}$, which is typically $O(m)$. While this is a somewhat more efficient computation, it is \emph{communication-intensive}, as it requires \emph{transferring all of the (either training or validation) data} at each FL round. Most importantly, it \emph{can not provide any privacy} to the users' data, an important, inherent requirement/constraint in FL. Finally, the loss function has to be strictly convex and twice differentiable (which is not always the case in modern ML applications). \cite{pmlrv70koh17a} proposed to swap out non-differentiable components for smoothed approximations, but there is no quality guarantee of the influence calculated in this way.

\subsection{Lazy Influence (LIA): A Practical Influence Metric for Filtering Data in FL Applications} \label{sec: proposed influence}

The key idea is that \emph{we do not need to approximate the influence value} to filter data; we only need an accurate estimate of its \emph{sign} (in expectation). Recall that a positive influence value indicates a data point improves model accuracy. Thus, we only need to approximate the sign of the loss and use that information to filter out data whose influence falls below a certain threshold.

Recall that each data holder participant may assume two roles: the role of the contributor ($A$) and the role of the validator ($B$). The proposed approach works as a `rite of passage' for contributor $A$, arriving at round $t$. Our approach works as follows (Algorithm \ref{algo:1}):

\textbf{\quad(i)} At federated learning round $t$ (model $M_t(\theta_t)$), the contributor participant $A$ performs a small number $k$ of local epochs to $M_t$ using a batch of his training data $Z_{A, t}$, resulting in $\tilde{\theta}_t^A$. $k$ is a hyperparameter. $\tilde{\theta}_t^A$ is the partially trained model of participant $A$, where most of the layers, except the last one, have been frozen. The model should not be fully trained for two key reasons: efficiency and avoiding over-fitting (e.g., in our simulations, we only performed 1-20 epochs). Furthermore, $A$ adds noise to $\tilde{\theta}_t^A$ (see Section \ref{sec: Differentially Private Sharing of the Partially Updated Joint Model}) to ensure strong, worst-case local differential privacy. Finally, $A$ sends only the last layer (to reduce communication cost) of  $\tilde{\theta}_t^A$ to every other participant.

\indent
\textbf{\quad(ii)} Each validator $B$ uses his validation dataset $Z_{val}^B$ to estimate the sign of the influence using Equation \ref{eq: proposed influence}. Next, the validator applies noise to $\mathcal{I}_{LIA}(Z_{val}^B)$, as described in Section \ref{sec: Differentially Private Reporting}, to ensure strong, worst-case differential privacy guarantees (i.e., keep his validation dataset private).
\small
\begin{equation} \label{eq: proposed influence}
    \mathcal{I}_{LIA}(Z_{val}^B) \triangleq \sign \left( \sum_{z_{val} \in Z_{val}^B}  L(z_{val}, \theta_t)  -  L(z_{val}, \theta_t^A) \right)
\end{equation}
\normalsize

\indent
\textbf{\quad(iii)} Finally, the center $C$ aggregates the obfuscated votes $\mathcal{I}_{LIA}(Z_{val}^B)$ from all validators and filters out data with cumulative score \emph{below a threshold} ($\sum_{\forall B} \mathcal{I}_{LIA}(Z_{val}^B) < T$). Specifically, we cluster the votes into two clusters (using k-means) and use the arithmetic mean of the cluster centers as the filtration threshold.

\begin{algorithm2e}[t]
  {
        
    \For{Data contributor $A$ arriving at round $t$}
    {
        
        $A$: Performs $k$ local epochs with $Z_{A,t}$ on the partially-frozen model $\tilde{\theta}_t^A$.
        
        $A$: Applies DP noise to $\tilde{\theta}_t^A$.
        
        $A$: sends last layer of $\tilde{\theta}_{t}^A$ to a subset of the participants, that will be acting as validators.
        
            \For{$B_i$ in $Validators(t)$}
            {
                $B_i$: Evaluates the loss of $Z_{val}^{B_i}$ on $\theta_t$ (current joint model of the federation)
                
                $B_i$: Evaluates the loss of $Z_{val}^B$ on $\tilde{\theta}_{t}^{A}$ (partially trained model from $A$)
                
                $B_i$: Calculates vote $v$ (Lazy Influence Approximation), according to Equation \ref{eq: proposed influence}
                
                $B_i$: Applies noise to $v$ according to his privacy parameter $p$ to get $v'$ (Equation \ref{eq: obfuscation})
            
                $B_i$: Sends $v'$ to $C$ (private majority voting)
            }
            
        $C$: Filters out (`rite of passage') $A$'s data if $\sum_{\forall B_i} \mathcal{I}_{LIA}(Z_{val}^{B_i}) < T$.
    }
  }    
  \caption{Filtering Poor Data Using the Lazy Influence Approximation (LIA) in FL}
  \label{algo:1}
\end{algorithm2e}

\subsubsection{Advantages of the proposed lazy influence}
Depending on the application, the designer may select any optimizer to perform the model updates. We do not require the loss function to be twice differentiable and convex; only once differentiable. It is significantly more \emph{computation and communication efficient}; an essential prerequisite for any FL application. This is because participant $A$ only needs to send (a \emph{small part} of) the model parameters $\theta$, not his training data. Moreover, computing a few model updates (using, e.g., SGD or any other optimizer) is significantly faster than computing either the exact influence or an approximation due to the numerous challenges. As a concrete example, computing the proposed LIA is \emph{~30 times faster} than the exact influence (on our simulations on real-data for Human Activity Recognition). Finally, and importantly, we ensure the \emph{privacy} of both the train and validation dataset of every participant.

\subsubsection{Sharing the Partially Updated Model: Privacy, Communication Cost, and Scalability}  \label{sec: Differentially Private Sharing of the Partially Updated Joint Model}

Each contributor $A$ shares a partially trained model $\tilde{\theta}_t^A$ (see step (i) of Section~\ref{sec:  proposed influence}). It is important to stress that A only sends the last layer of the model. This has two significant benefits. First, it minimizes the impact of the differential privacy noise. We follow ~\cite{abadi2016deep, opacus} to ensure strong datapoint-level local differential privacy guarantees by (i) imposing a bound on the gradient (using a clipping threshold $\Delta$), and (ii) adding carefully crafted Gaussian noise (parameterized by $\sigma$). Second, it \emph{reduces the communication overhead} (e.g., in our CIFAR simulations, \emph{we only send $0.009\%$ of the model's weights}). Moreover, as explained in the Introduction, this communication cost will be incurred only \emph{once}, when we use our approach as a `right of passage’ every time a participant joins the federation. Importantly, in practice one can randomly select a (sufficiently large\footnote{With enough votes, the aggregated LIA signs will match the sign of the true influence value in expectation. The size of the validators set is problem specific. See Sections \ref{sec: Evaluation Results}, \ref{sec: Privacy}, and Figure \ref{fig:non-iid epsilon}.}) \emph{subset} of validators, keeping the communication cost \emph{constant} as we scale.

\begin{figure*}[!ht]
\centering
\begin{subfigure}{.33\textwidth}
  \centering
  \includegraphics[width=1\linewidth, height=9em, clip]{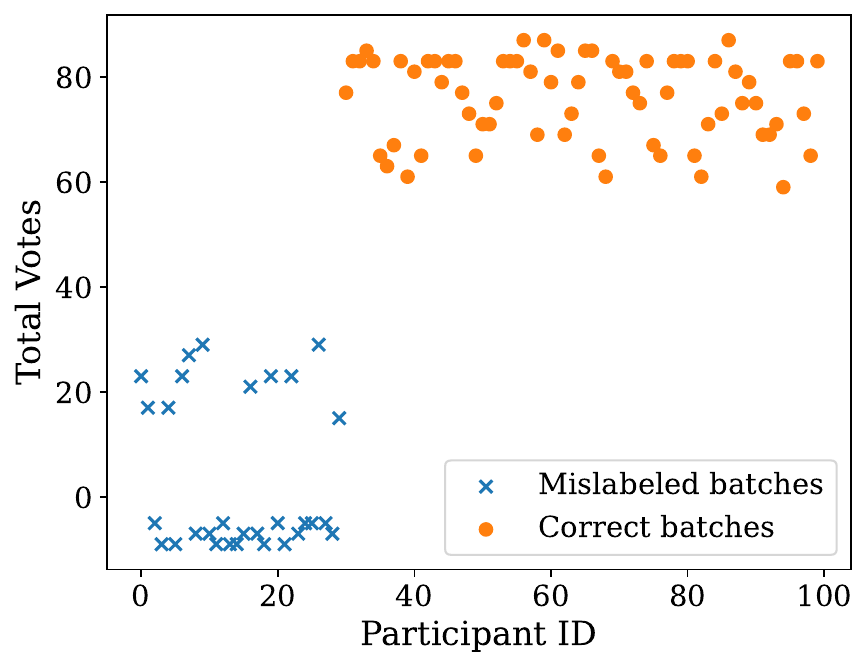}
  \caption{ }
  \label{fig:sub1}
\end{subfigure}%
\begin{subfigure}{.33\textwidth}
  \centering
  \includegraphics[width=1\linewidth, height=9em, clip]{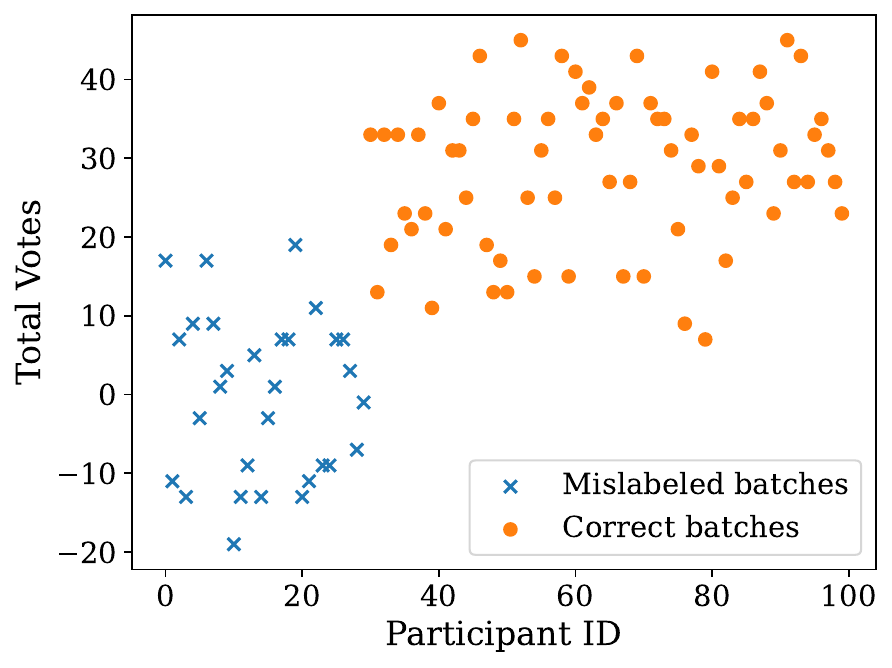}
  \caption{ }
  \label{fig:sub2}
\end{subfigure}
\begin{subfigure}{.33\textwidth}
  \centering
  \includegraphics[width=1\linewidth, height=9em, clip]{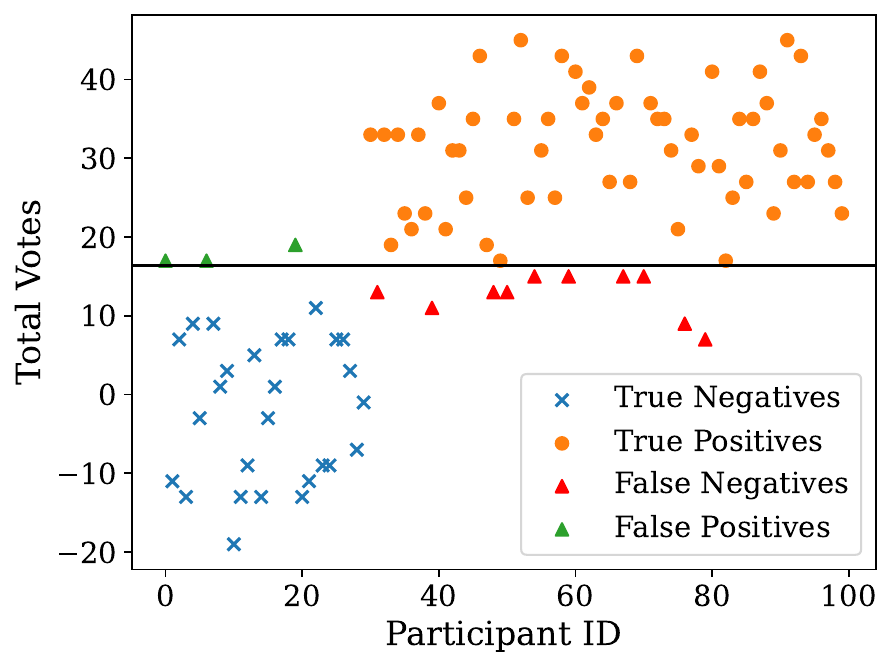}
  \caption{ }
  \label{fig:sub3}
\end{subfigure}
\caption{Visualization of the private voting scheme. The $x$-axis represents a contributor participant $A$. The $y$-axis shows the sum of all votes from all the validators, i.e., $\sum_{\forall B} \mathcal{I}_{LIA}(Z_{val}^B)$. Figure~\ref{fig:sub1} corresponds to the sum of true votes (no privacy) for the valiation data of each contributor on the $x$-axis, while Figure~\ref{fig:sub2} depicts the sum of differentially private votes ($\varepsilon=1$), according to randomized reporting algorithm. Finally, Figure~\ref{fig:sub3} shows the filtration threshold, corresponding to the arithmetic mean of the two cluster centers (computed using k-means).}
\label{fig:dp voting}
\end{figure*}

\subsubsection{Differentially Private Reporting of the Influence}  \label{sec: Differentially Private Reporting}

Along with the training data, we also need to ensure the privacy of the validation data used to calculate the influence. Protecting the validation data in an FL setting is critical since (i) it is an important constraint of the FL setting, (ii) participants want to keep their sensitive information (and potential means of income, e.g., in a crowdsourcing application) private, and (iii) the center wants to ensure that malicious participants can not tailor their contributions to the validation sets.

Each validator $B_i$ is assigned a random ID number every time they act in that capacity. Moreover, each can split their test data into multiple \emph{disjoint} sets, and then use a each set \emph{only once}.\footnote{E.g., in our applications, each of these sets required at most 50 points for the vote to be accurate.} We obfuscate their influence reports using RAPPOR~\cite{erlingsson2014rappor}, which results in an $\varepsilon$-differential privacy guarantee~\cite{dwork2006calibrating}.\footnote{Note that RAPPOR provides significantly stronger privacy guarantees ($\delta=0$) that what is normally deployed in modern systems. Thus, in practice, the privacy loss $\varepsilon$ will be even lower if we allow $\delta >0$.} The obfuscation (permanent randomized response~\cite{warner1965randomized}) takes as input the participant's true influence value $v$ (binary) and privacy parameter $p$, and creates an obfuscated (noisy) reporting value $v'$, according to Eq. \ref{eq: obfuscation}. 
$p$ is a \emph{user-tunable} parameter that allows the participants themselves to \emph{choose their desired level of privacy}, while maintaining reliable filtering. The worst-case privacy guarantee can be computed by each participant \emph{a priori}, using Eq. \ref{eq: dp guarantee}~\cite{erlingsson2014rappor}.
\small
\begin{equation} \label{eq: obfuscation}
    v' = \begin{cases}
    +1, & \text{with probability $\frac{1}{2}p$} \\
    -1, & \text{with probability $\frac{1}{2}p$} \\
    v, & \text{with probability $1 - p$}
    \end{cases}
\end{equation}
\begin{equation} \label{eq: dp guarantee}
    \varepsilon = 2\ln\left(\frac{1 - \frac{1}{2}p}{\frac{1}{2}p}\right)
\end{equation}
\normalsize

It is important to note that in a Federated Learning application, the center $C$ aggregates the influence sign from  \emph{a large number of participants}. This means that even under really strict privacy guarantees, \emph{the aggregated influence signs (which is exactly what we use for filtering) will match the sign of the true value} in expectation. This results in high-quality filtering, as we will demonstrate in Section \ref{sec: Evaluation Results}.

We visualize the obfuscation process in Figure~\ref{fig:dp voting}, which
shows the sum of true votes ($y$-axis) for the validation data of each contributor ($x$-axis). Here we can see a clear distinction in votes between corrupted and correct batches. Most of the corrupted batches (corrupted contributor participants) take negative values, meaning that the majority of the validators voted against them. In contrast, the correct batches are close to the upper bound.  Figure~\ref{fig:sub2} demonstrates the effect of applying DP noise ($\varepsilon=1$) to the votes: differentiating between the two groups becomes more challenging. To find an effective decision threshold, we use k-means to cluster the votes into two clusters and use the arithmetic mean of the cluster centers as the filtration threshold (Figure~\ref{fig:sub3}).

\section{Evaluation Results} \label{sec: Evaluation Results}

\paragraph{Dataset}  \label{sec: Dataset}

We evaluated the proposed approach on four well-established datasets: \emph{CIFAR10}~\cite{cifar10}, \emph{CIFAR100}~\cite{cifar10}, \emph{real-data} on Human Activity Recognition (HAR) on edge devices~\cite{anguita2013public,HAR_dataset_kaggle}, and \emph{real-data} on diagnosing heart conditions from X-rays~\cite{rajpurkar2017chexnet}. 
Between these datasets, we evaluate on \textbf{two different modalities} (image data and sensor data), and \textbf{corruption of both the input X, and the labels y}. Due to lack of space, we focus on CIFAR in this section. Similar results were achieved for the other datasets. Please see the supplement (\ref{sec: numerical results}) for detailed results.

\paragraph{Corruption Methods} 
(i) \textbf{Random label}: a random label is sampled for every training point. Used for the IID setting (as it does not make sense to assign a random label to a highly skewed Non-IID setting). (ii) \textbf{Label shift}: Every correct label is mapped to a different label and this new mapping is applied to the whole training dataset. Used in both IID and non-IID settings.

\paragraph{Setup}  \label{sec: Setup}

Our evaluation involves a single round of Federated Learning. A small portion of every dataset (around 1\%) is selected as the `warm-up' data used by the center $C$ to train the initial model $M_0$. Each participant has two datasets: a training batch ($Z_A$, see Section \ref{sec: proposed influence}, step (i)), which the participant uses to update the model when acting as the contributor participant, and a validation dataset ($Z_{val}^B$, see Section \ref{sec: proposed influence}, step (ii)), which the participant uses to estimate the sign of the influence when acting as a validator participant. 
As a concrete example, for the CIFAR simulations the ratio of these datasets is $2:1$. The training batch size is 100 (i.e., the training dataset includes 100 points, and the validation dataset consists of 50 points). This means that, e.g., for a simulation with 100 participants, each training batch is evaluated on $50 \times (100-1)$ validation points, and that for each training batch (contributor participant $A$), the center collected ($100-1$) estimates of the influence sign (Equation \ref{eq: proposed influence}). See the supplement~(\ref{sec: numerical results}) for details on the HAR dataset.
We report results when having 30\% corrupt participants. Additional results (0\%, 10\%, 20\%, 40\%) can be found in the supplement~\ref{sec: numerical results}. For each corrupted batch, we corrupted 100\% of the data points (similar results were achieved with 90\%). Each simulation was run \emph{8 times}. We report average values and standard deviations. The proposed approach is \textbf{model-agnostic} and can be used with \emph{any} gradient-descent-based ML method.

\paragraph{Non-IID Setting}  \label{sec: Non-IID Setting}

The main hurdle for FL is that not all data is IID. Heterogeneous data distributions are all but uncommon in the real world. To simulate non-IID data, we used the Dirichlet distribution to split the training dataset as in related literature~\cite{hsu2019measuring,lin2020ensemble,hoech2022,yu2022tct}. This distribution is parameterized by $\alpha$, which controls the concentration of different classes (see the supplement~\ref{sec: Appendix-Non-IID Setting} for a visualization). We report results for $\alpha \rightarrow 0.1$, as in related literature (e.g., \cite{yu2022tct}), see the supplement~\ref{sec: numerical results} for the rest. This translates to a \emph{highly non-IID distribution}, e.g., with just 3 classes per participant for the HAR dataset.

\paragraph{Baselines} 

We compare against four baselines. \textbf{(i) Corrupted model}: no sanitization (no filtering). \textbf{(ii) Oracle filtration}: ideal scenario where we know which participants contribute bad data. \textbf{(iii) KRUM}~\cite{KRUM}: byzantine robustness technique that selects the best gradient out of the update based on a pair-wise distance metric. \textbf{(iv) Trimmed-mean}~\cite{pmlr-v80-yin18a}: another byzantine robustness technique that takes the average of gradients instead of just selecting one, also based on a pair-wise distance metric. \textbf{(v) Centered-Clipping}~\cite{karimireddy2020learning}: state-of-the-art technique for byzantine robust aggregators. Importantly, we show that our filtration technique is not mutually exclusive with these aggregators, instead it is highly beneficial \emph{in conjunction} to them (see Figure~\ref{fig:performance}).

\subsection{Model Performance}  \label{sec: Model Accuracy}

\begin{figure*}[t!]
\centering
\begin{subfigure}{.35\textwidth}
  \centering
  \includegraphics[width=\linewidth]{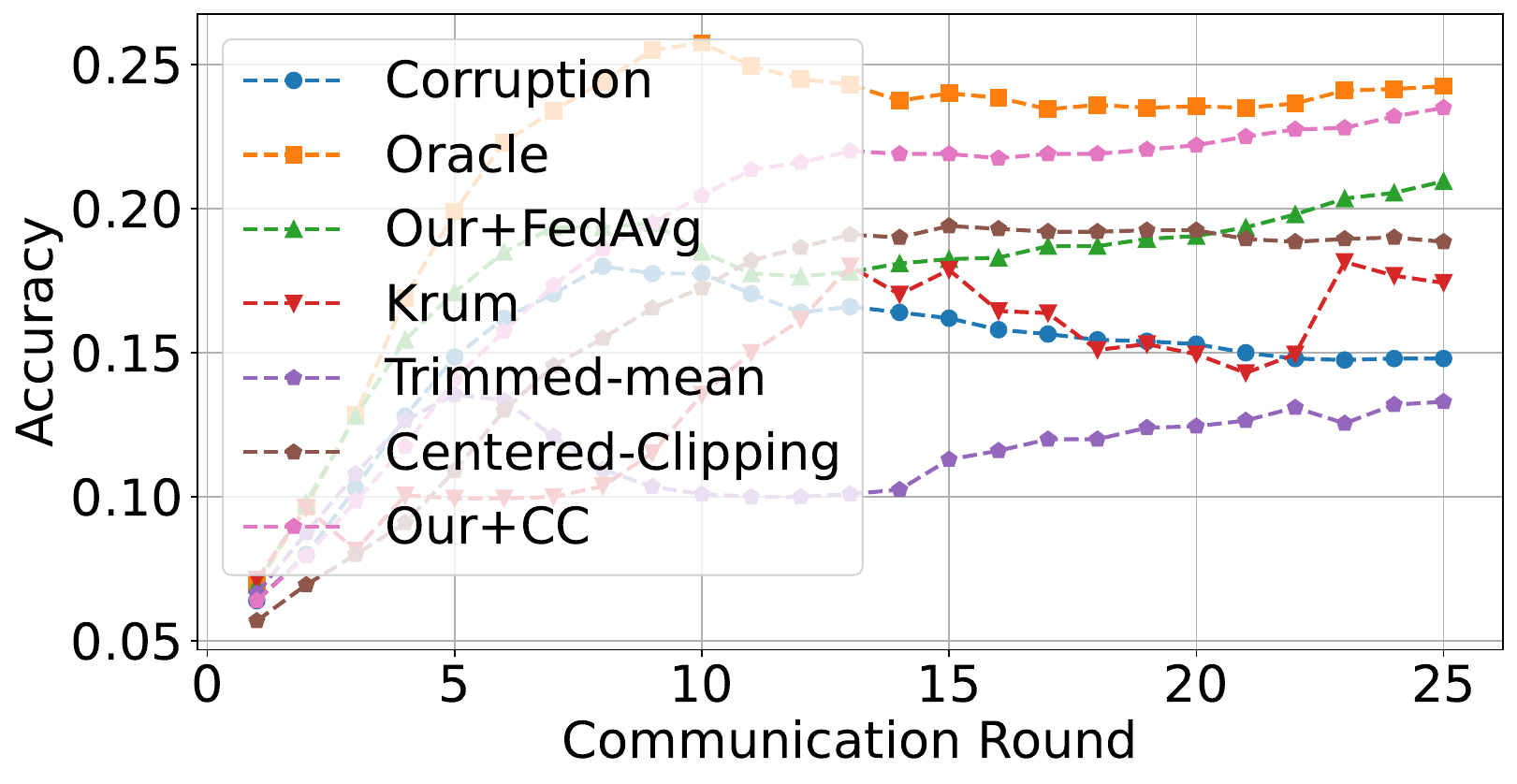}
  \caption{Label shift Non-IID ($\alpha = 0.01$)}
    \label{fig:random_label}
\end{subfigure}%
\begin{subfigure}{.35\textwidth}
  \centering
  \includegraphics[width=\linewidth]{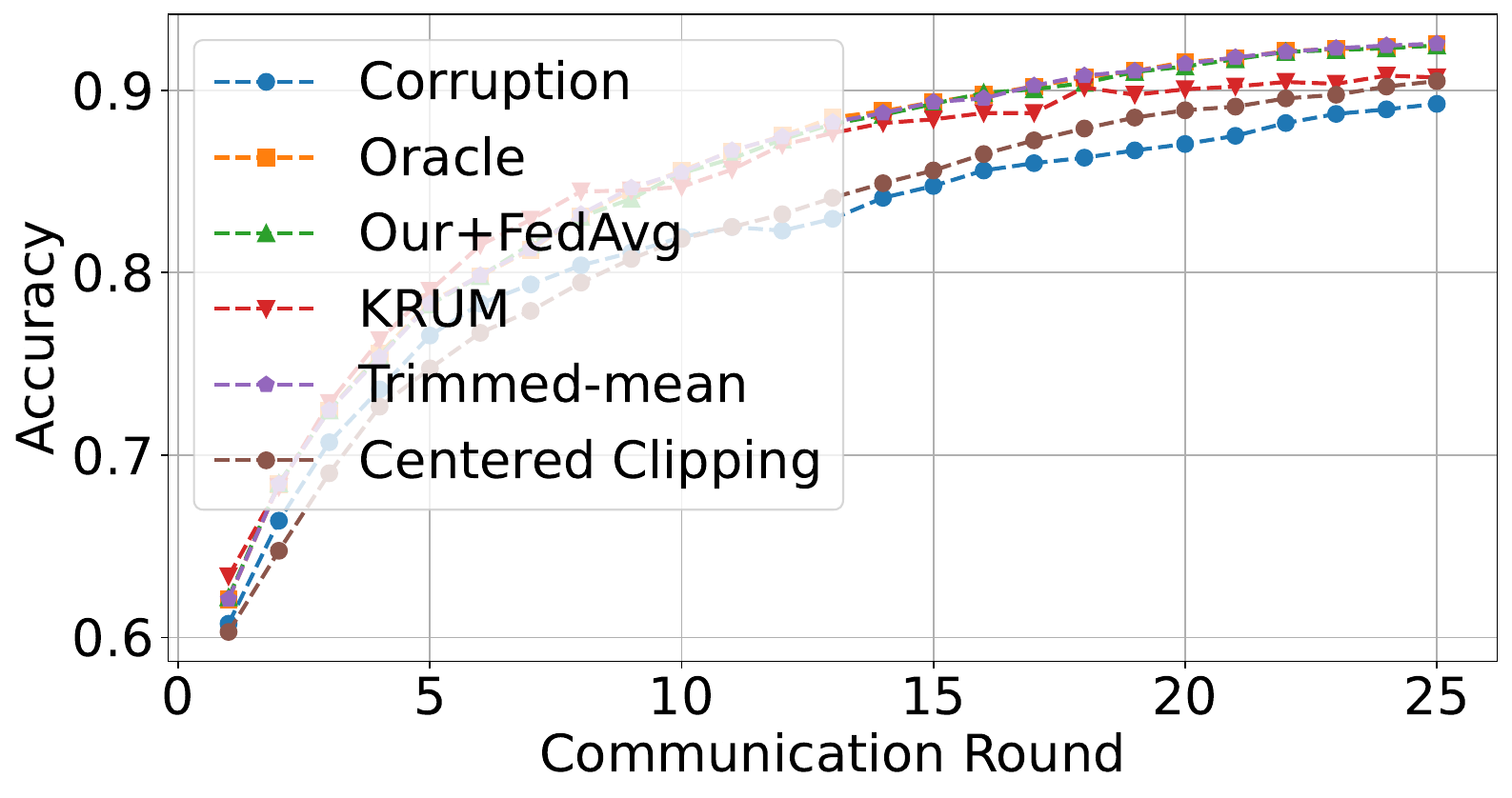}
  \caption{Random Label IID}
    \label{fig:random_label_iid}
\end{subfigure}
\caption{Model performance (model accuracy) over 25 communication rounds. 30\% mislabel rate on CIFAR-10. The proposed (LIA) and oracle filters are used only \emph{once} at the start (`right of passage' scenario). We compare a centralized model with no filtering (\textcolor{figure_blue}{blue}) to an FL model under perfect (oracle) filtering (\textcolor{figure_orange}{orange}),  KRUM (\textcolor{figure_red}{red}), Trimmed-mean (\textcolor{figure_purple}{purple}), Centered-Clipping(\textcolor{figure_brown}{brown}), our approach with FedAvg (\textcolor{figure_green}{green}), and our approach with Centered-Clipping (\textcolor{figure_pink}{pink}). Note that the jagged line for KRUM is because only a single gradient is selected instead of performing FedAvg.}
\label{fig:performance}
\end{figure*}

The proposed approach achieves \emph{high model performance, close to the model performance of the perfect (oracle) filtering} ($13.6\%$ worse in the non-IID setting, and $0.1\%$ in the IID setting). Focusing on the non-IID setting (Figure \ref{fig:random_label}), which is the more challenging and relevant for FL, our approach achieves a $10.8\%$, $20.3\%$, and $57.5\%$ improvement over the Centered-Clipping, KRUM, and Trimmed-mean baselines, respectively, after 25 communication rounds. Interestingly, combining our approach with Centered-Clipping provides results almost as good as the Oracle filter (only $3\%$ worse), which comes to show that LIA can \emph{complement} Byzantine robustness techniques. Finally, while in the IID setting all methods perform similarly (Figure \ref{fig:random_label_iid}), recall that the baselines do not provide privacy guarantees.

\begin{figure*}[t!]
\centering
\begin{subfigure}{.35\textwidth}
  \centering
  \includegraphics[width=\linewidth]{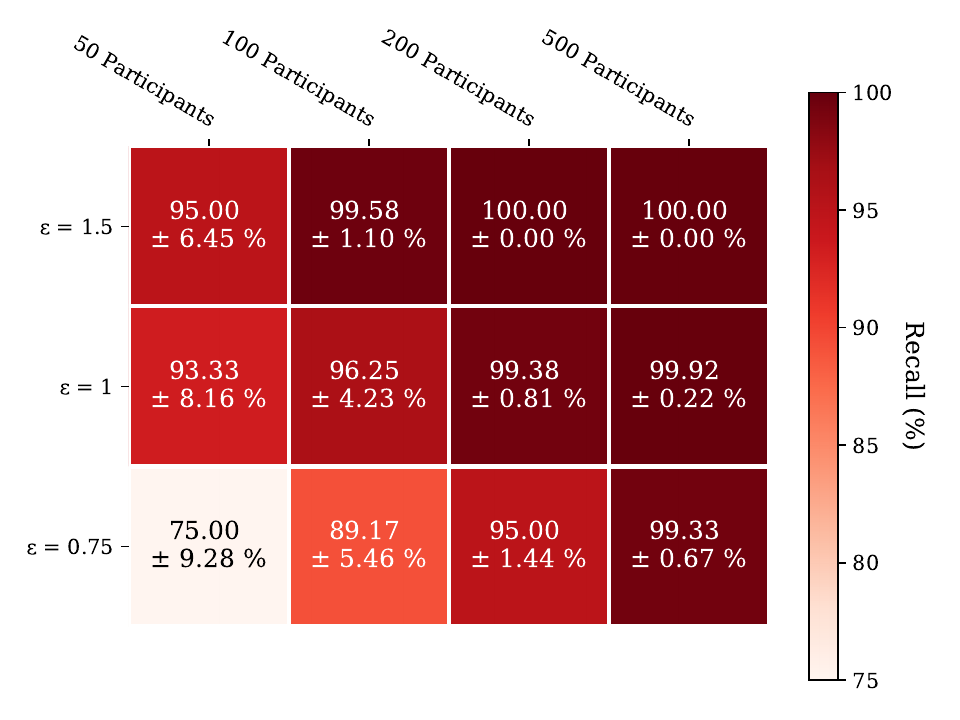}
  \caption{Recall}
    \label{fig:non-iid Recall}
\end{subfigure}%
\begin{subfigure}{.35\textwidth}
  \centering
  \includegraphics[width=\linewidth, clip]{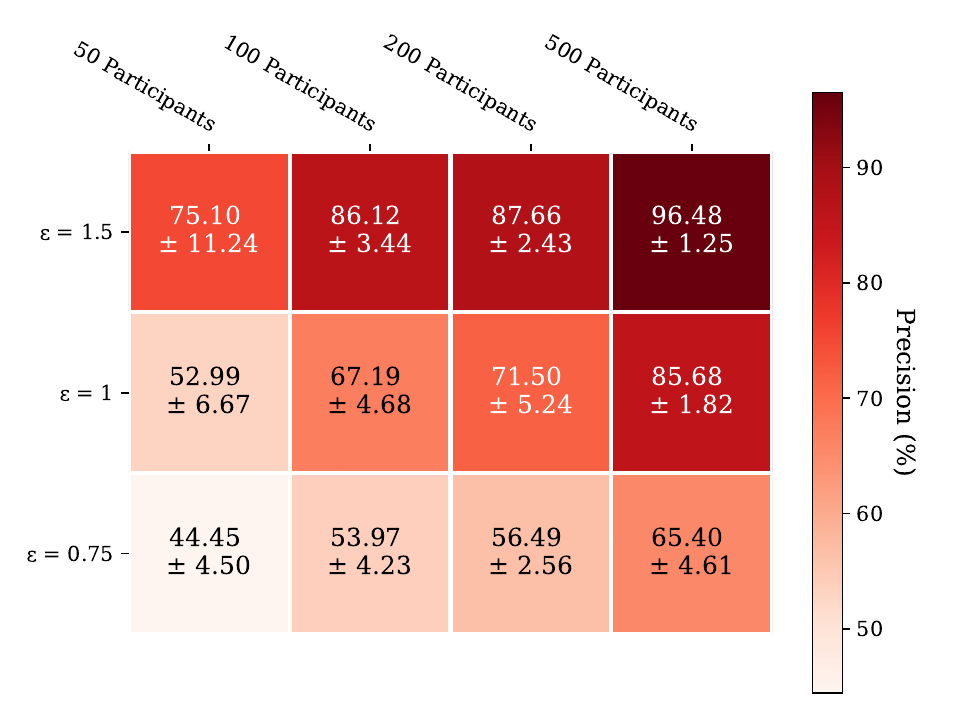}
  \caption{Precision}
    \label{fig:non-iid Precision}
\end{subfigure}
\caption{Recall and Precision on CIFAR 10, highly non-IID ($\alpha \rightarrow 0.1$), for increasing problem size (\# of participants), and varying privacy guarantees (lower $\varepsilon$ provides stronger privacy). $\delta = 10^{-5}$.}\label{fig:non-iid epsilon}
\end{figure*}

\subsection{Filtration Metrics: Recall, Precision, Accuracy}  \label{sec: Precision and Recall}

Recall is the most informative metric to evaluate the efficiency of our filtering approach. Recall refers to the ratio of detected mislabeled batches over all of the mislabeled batches. \emph{Including a mislabeled batch can harm a model's performance significantly more compared to removing an unaltered batch.} Thus, achieving \emph{high recall} is of paramount importance. Meanwhile, precision represents the ratio of correctly identified mislabeled batches over all batches identified as mislabeled. An additional benefit of using the proposed lazy influence metric for scoring data is that it also allows us to identify correctly labeled data, which nevertheless do not provide a significant contribution to the model. Finally, filtration accuracy refers to the ratio of correctly identified batches over all batches.

The proposed approach achieves both high recall and precision (Figure \ref{fig:non-iid epsilon}), despite the \emph{high degree of non-IID} (low concentration of classes per participant). Notably, the metrics \emph{improve significantly as we increase the number of participants} (horizontal axis, Figure \ref{fig:non-iid epsilon}). In simple terms, more validators mean more samples of the different distributions. Thus, `honest' participants get over the filtering threshold, even in highly non-IID settings. Recall reaches $100\%$ and precision $96.48\%$ on CIFAR10 (Figure \ref{fig:non-iid epsilon})
by increasing the number of participants to just 500 (highly non-IID setting and under really strict worst-case privacy guarantees).
Results for the IID setting are significantly better, with almost \emph{perfect} recall, precision, and accuracy, even with only 100 participants. 

\subsection{Privacy}  \label{sec: Privacy}

As expected, there is a trade-off between privacy and filtration quality (see Figure~\ref{fig:non-iid epsilon}, vertical axis, where $\varepsilon$ refers to the privacy guarantee for both the training and validation data/participant votes). Nevertheless, Figure~\ref{fig:non-iid epsilon} demonstrates that our approach can provide \emph{reliable filtration}, even under \emph{really strict, worst-case privacy requirements} ($\varepsilon = 1$, which is the recommended value in the DP literature~\cite{triastcyn2020data}, $\delta = 10^{-5}$). Importantly, our decentralized framework allows each participant to \emph{compute} and \emph{tune} his \emph{own} worst-case privacy guarantee \emph{a priori} (see Section \ref{sec: Differentially Private Reporting}).

The \emph{privacy trade-off can be mitigated}, and the quality of the filtration can be significantly improved by increasing the number of validators (Figure~\ref{fig:non-iid epsilon}, horizontal axis). The higher the number of validators, the better the filtration (given a fixed number of corrupted participants). This is because as the number of validators increases, the aggregated influence signs (precisely what we use for filtering) will match the sign of the true influence value (in expectation). For $500$ validators, we achieve high-quality filtration even for $\varepsilon = 0.75$. This is important given that in most real-world FL applications, we \emph{expect a large number of validators}.

\subsection{Robustness}

We evaluated our approach for varying mislabeling percentage (0\%, 10\%, 20\%, 30\%, 40\%), varying levels of non-IID-ness ($\alpha = 0.1, 1, 10$), and a setting where not all data in a mislabeled batch are corrupt (90\% corrupt). In all cases, the proposed approach enabled effective filtering (recall of $>90\%$). Please see the supplement~\ref{sec: Appendix-Non-IID Setting} for detailed results.

\subsection{Minority Distributions} \label{sec: Minority Distributions}

An important concern when filtering data is to ensure that clients with minority distributions will not be filtered out. Our results thus far demonstrate that our proposed approach achieves high quality filtering (outperforming established baselines), even on very hard non-IID settings (as low as just 3 classes per participant). Yet, these are aggregate results. To investigate specifically the case where clients have data from minority distributions, we consider the most \emph{extreme} scenario, where \emph{only one client has data from a specific class} (consider e.g., a hospital that treats a rare disease). 

\begin{figure}
    \centering
    \includegraphics[width=\linewidth, clip, trim={0em 0em 0em 0em}]{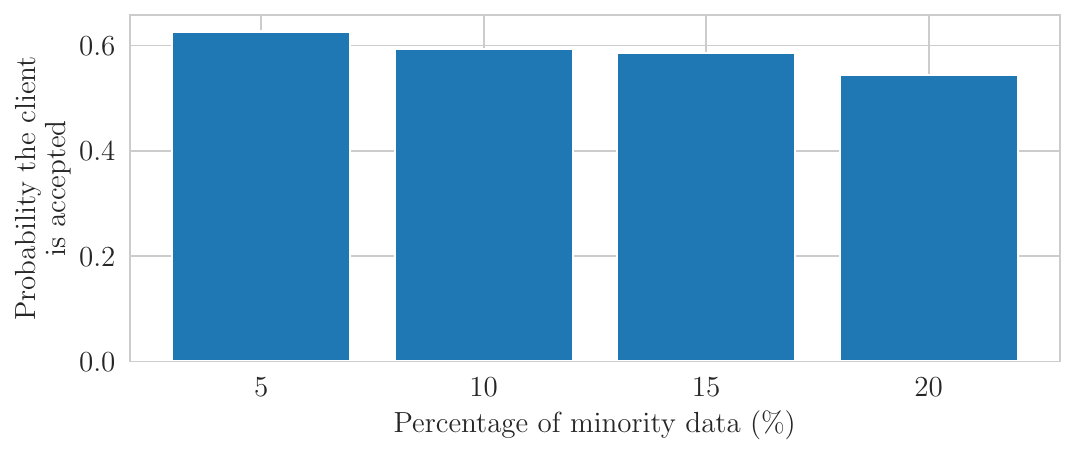}
    \caption{Probability that a client with data from a unique distribution (see Section \ref{sec: Minority Distributions}) will be accepted in the federation. With high probability ($> 50\%$) the client will \emph{not} be filtered out, even in the extreme case of having 20\% of data from a unique distribution. Results averaged over 512 runs.}
    \label{figure:minority-distributions}
\end{figure}

Specifically, we consider a scenario where every validator has data from all but one classes (leave-one-out, i.e., $\mathcal{Y}\setminus \{ y_{out} \}$), drawn from a Dirichlet distribution, in a highly non-IID setting (as in our other simulations). Let a new incoming client $A$ have data from all class (including $y_{out}$). Let 5\%, 10\%, 15\%, 20\% of client $A$'s data belong to his unique class $y_{out}$, while the rest are drawn from classes $\mathcal{Y}\setminus \{ y_{out} \}$ using the Dirichlet distribution. Figure \ref{figure:minority-distributions} shows that, with high probability ($0.63$ for 5\%, $0.54$ for 20\%), client $A$ will \emph{not} be filtered out. Our intuition is that client $A$ also has data that (partially) overlap with the validators (even in highly non-IID settings), and this overlap results in positive influence, in spite of their minority class data.

\section{Conclusion}  \label{sec: Conclusion}

The quality of a model obtained through Federated Learning can only be as good as the quality of the contributed data.
Mislabeled, corrupted, or even malicious data can result in a strong degradation of the performance of the model (Figure~\ref{figure:1}).
However, it is challenging to ensure this quality while preserving the privacy guarantees that are a hallmark of federated learning, and at the same time ensure that data from minority classes is not discarded.
We propose a \emph{practical} influence approximation (\emph{`lazy influence'}) to be used as a `rite of passage',
allowing for effective filtering (recall of $>90\%$, and even up to 100\%), while providing \emph{strict, worst-case} $\varepsilon$-differential privacy guarantees ($\varepsilon \leq 1$) for both the training and validation data.

\bibliographystyle{plain}
\bibliography{IEEEBigData2024}

\begin{thebibliography}{10}

\bibitem{abadi2016deep}
Martin Abadi, Andy Chu, Ian Goodfellow, H~Brendan McMahan, Ilya Mironov, Kunal
  Talwar, and Li~Zhang.
\newblock Deep learning with differential privacy.
\newblock In {\em CSS 2016 ACM SIGSAC}, 2016.

\bibitem{HAR_dataset_kaggle}
Davide Anguita, Alessandro Ghio, Luca Oneto, Xavier Parra, Jorge~Luis
  Reyes-Ortiz, et~al.
\newblock Human activity recognition with smartphones dataset, 2013.
\newblock
  \url{https://www.kaggle.com/datasets/uciml/human-activity-recognition-with-smartphones}.

\bibitem{anguita2013public}
Davide Anguita, Alessandro Ghio, Luca Oneto, Xavier Parra, Jorge~Luis
  Reyes-Ortiz, et~al.
\newblock A public domain dataset for human activity recognition using
  smartphones.
\newblock In {\em Esann}, volume~3, page~3, 2013.

\bibitem{augenstein2019generative}
Sean Augenstein, H~Brendan McMahan, Daniel Ramage, Swaroop Ramaswamy, Peter
  Kairouz, Mingqing Chen, Rajiv Mathews, et~al.
\newblock Generative models for effective ml on private, decentralized
  datasets.
\newblock {\em arXiv preprint arXiv:1911.06679}, 2019.

\bibitem{KRUM}
Peva Blanchard, El~Mahdi El~Mhamdi, Rachid Guerraoui, and Julien Stainer.
\newblock Machine learning with adversaries: Byzantine tolerant gradient
  descent.
\newblock In I.~Guyon, U.~Von Luxburg, S.~Bengio, H.~Wallach, R.~Fergus,
  S.~Vishwanathan, and R.~Garnett, editors, {\em Advances in Neural Information
  Processing Systems}, 2017.

\bibitem{caldas2018leaf}
Sebastian Caldas, Sai Meher~Karthik Duddu, Peter Wu, Tian Li, Jakub
  Kone{\v{c}}n{\`y}, H~Brendan McMahan, Virginia Smith, and Ameet Talwalkar.
\newblock Leaf: A benchmark for federated settings.
\newblock {\em arXiv preprint arXiv:1812.01097}, 2018.

\bibitem{cao2020fltrust}
Xiaoyu Cao, Minghong Fang, Jia Liu, and Neil~Zhenqiang Gong.
\newblock Fltrust: Byzantine-robust federated learning via trust bootstrapping.
\newblock {\em arXiv:2012.13995}, 2020.

\bibitem{chen2019federated}
Mingqing Chen, Rajiv Mathews, Tom Ouyang, and Fran{\c{c}}oise Beaufays.
\newblock Federated learning of out-of-vocabulary words.
\newblock {\em arXiv preprint arXiv:1903.10635}, 2019.

\bibitem{choquette2022multi}
Christopher~A Choquette-Choo, H~Brendan McMahan, Keith Rush, and Abhradeep
  Thakurta.
\newblock Multi-epoch matrix factorization mechanisms for private machine
  learning.
\newblock {\em arXiv preprint arXiv:2211.06530}, 2022.

\bibitem{christmann2004robustness}
Andreas Christmann and Ingo Steinwart.
\newblock On robustness properties of convex risk minimization methods for
  pattern recognition.
\newblock {\em JMLR}, 2004.

\bibitem{cook1980characterizations}
R~Dennis Cook and Sanford Weisberg.
\newblock Characterizations of an empirical influence function for detecting
  influential cases in regression.
\newblock {\em Technometrics}, 1980.

\bibitem{cook1982residuals}
R~Dennis Cook and Sanford Weisberg.
\newblock {\em Residuals and influence in regression}.
\newblock New York: Chapman and Hall, 1982.

\bibitem{DanassisPalma}
Panayiotis Danassis, Aleksei Triastcyn, and Boi Faltings.
\newblock A distributed differentially private algorithm for resource
  allocation in unboundedly large settings.
\newblock In {\em Proceedings of the 21th International Conference on
  Autonomous Agents and MultiAgent Systems, {AAMAS-22}}. International
  Foundation for Autonomous Agents and Multiagent Systems, 2022.

\bibitem{dasgupta2009sampling}
Anirban Dasgupta, Petros Drineas, Boulos Harb, Ravi Kumar, and Michael~W
  Mahoney.
\newblock Sampling algorithms and coresets for $\backslash$ell\_p regression.
\newblock {\em SIAM Journal on Computing}, 2009.

\bibitem{de2204unlocking}
Soham De, Leonard Berrada, Jamie Hayes, Samuel~L Smith, and Borja Balle.
\newblock Unlocking high-accuracy differentially private image classification
  through scale, 2022.
\newblock {\em arXiv:2204.13650}, 2022.

\bibitem{deng2009imagenet}
Jia Deng, Wei Dong, Richard Socher, Li-Jia Li, Kai Li, and Li~Fei-Fei.
\newblock Imagenet: A large-scale hierarchical image database.
\newblock In {\em 2009 IEEE conference on computer vision and pattern
  recognition}, pages 248--255. Ieee, 2009.

\bibitem{dwork2006}
Cynthia Dwork.
\newblock Differential privacy.
\newblock In {\em 33rd International Colloquium on Automata, Languages and
  Programming, part II (ICALP 2006)}, volume 4052, pages 1--12, Venice, Italy,
  July 2006. Springer Verlag.

\bibitem{10.1007/11787006_1}
Cynthia Dwork.
\newblock Differential privacy.
\newblock In Michele Bugliesi, Bart Preneel, Vladimiro Sassone, and Ingo
  Wegener, editors, {\em Automata, Languages and Programming}, pages 1--12,
  Berlin, Heidelberg, 2006. Springer Berlin Heidelberg.

\bibitem{dwork2006our}
Cynthia Dwork, Krishnaram Kenthapadi, Frank McSherry, Ilya Mironov, and Moni
  Naor.
\newblock Our data, ourselves: Privacy via distributed noise generation.
\newblock In {\em Annual International Conference on the Theory and
  Applications of Cryptographic Techniques}, pages 486--503. Springer, 2006.

\bibitem{dwork2006calibrating}
Cynthia Dwork, Frank McSherry, Kobbi Nissim, and Adam Smith.
\newblock Calibrating noise to sensitivity in private data analysis.
\newblock In {\em Theory of cryptography conference}, 2006.

\bibitem{dwork2014algorithmic}
Cynthia Dwork, Aaron Roth, et~al.
\newblock The algorithmic foundations of differential privacy.
\newblock {\em Foundations and Trends in Theoretical Computer Science},
  9(3-4):211--407, 2014.

\bibitem{erlingsson2014rappor}
{\'U}lfar Erlingsson, Vasyl Pihur, and Aleksandra Korolova.
\newblock Rappor: Randomized aggregatable privacy-preserving ordinal response.
\newblock In {\em Proceedings of the 2014 ACM SIGSAC conference on computer and
  communications security}, pages 1054--1067, 2014.

\bibitem{ghorbani2019data}
Amirata Ghorbani and James Zou.
\newblock Data shapley: Equitable valuation of data for machine learning.
\newblock In {\em International Conference on Machine Learning}, pages
  2242--2251. PMLR, 2019.

\bibitem{pmlr-v97-ghorbani19c}
Amirata Ghorbani and James Zou.
\newblock Data shapley: Equitable valuation of data for machine learning.
\newblock In Kamalika Chaudhuri and Ruslan Salakhutdinov, editors, {\em
  Proceedings of the 36th International Conference on Machine Learning},
  volume~97 of {\em Proceedings of Machine Learning Research}, pages
  2242--2251. PMLR, 09--15 Jun 2019.

\bibitem{hard2018federated}
Andrew Hard, Kanishka Rao, Rajiv Mathews, Swaroop Ramaswamy, Fran{\c{c}}oise
  Beaufays, Sean Augenstein, Hubert Eichner, Chlo{\'e} Kiddon, and Daniel
  Ramage.
\newblock Federated learning for mobile keyboard prediction.
\newblock {\em arXiv preprint arXiv:1811.03604}, 2018.

\bibitem{he2020fedml}
Chaoyang He, Songze Li, Jinhyun So, Xiao Zeng, Mi~Zhang, Hongyi Wang, Xiaoyang
  Wang, Praneeth Vepakomma, Abhishek Singh, Hang Qiu, et~al.
\newblock Fedml: A research library and benchmark for federated machine
  learning.
\newblock {\em arXiv preprint arXiv:2007.13518}, 2020.

\bibitem{hoech2022}
Haley Hoech, Roman Rischke, Karsten Müller, and Wojciech Samek.
\newblock Fedauxfdp: Differentially private one-shot federated distillation,
  2022.

\bibitem{hsu2019measuring}
Tzu-Ming~Harry Hsu, Hang Qi, and Matthew Brown.
\newblock Measuring the effects of non-identical data distribution for
  federated visual classification.
\newblock {\em arXiv preprint arXiv:1909.06335}, 2019.

\bibitem{jia2019efficient}
Ruoxi Jia, David Dao, Boxin Wang, Frances~Ann Hubis, Nezihe~Merve Gurel, Bo~Li,
  Ce~Zhang, Costas~J Spanos, and Dawn Song.
\newblock Efficient task-specific data valuation for nearest neighbor
  algorithms.
\newblock {\em arXiv preprint arXiv:1908.08619}, 2019.

\bibitem{jia2019towards}
Ruoxi Jia, David Dao, Boxin Wang, Frances~Ann Hubis, Nick Hynes, Nezihe~Merve
  Gurel, Bo~Li, Ce~Zhang, Dawn Song, and Costas Spanos.
\newblock Towards efficient data valuation based on the shapley value.
\newblock In {\em AISTATS}, 2019.

\bibitem{kairouz2021practical}
Peter Kairouz, Brendan McMahan, Shuang Song, Om~Thakkar, Abhradeep Thakurta,
  and Zheng Xu.
\newblock Practical and private (deep) learning without sampling or shuffling.
\newblock In {\em International Conference on Machine Learning}, pages
  5213--5225. PMLR, 2021.

\bibitem{kairouz2021advances}
Peter Kairouz, H~Brendan McMahan, Brendan Avent, Aur{\'e}lien Bellet, Mehdi
  Bennis, Arjun~Nitin Bhagoji, Kallista Bonawitz, Zachary Charles, Graham
  Cormode, Rachel Cummings, et~al.
\newblock Advances and open problems in federated learning.
\newblock {\em Foundations and Trends{\textregistered} in Machine Learning},
  14(1--2):1--210, 2021.

\bibitem{karimireddy2020learning}
Sai~Praneeth Karimireddy, Lie He, and Martin Jaggi.
\newblock {Learning from History for Byzantine Robust Optimization}.
\newblock In {\em ICML 2021 - Proceedings of International Conference on
  Machine Learning}, 2021.

\bibitem{kasiviswanathan2011can}
Shiva~Prasad Kasiviswanathan, Homin~K Lee, Kobbi Nissim, Sofya Raskhodnikova,
  and Adam Smith.
\newblock What can we learn privately?
\newblock {\em SIAM Journal on Computing}, 40(3):793--826, 2011.

\bibitem{pmlrv70koh17a}
Pang-Wei Koh and Percy Liang.
\newblock Understanding black-box predictions via influence functions.
\newblock In {\em International conference on machine learning}, pages
  1885--1894. PMLR, 2017.

\bibitem{cifar10}
Alex Krizhevsky, Geoffrey Hinton, et~al.
\newblock Learning multiple layers of features from tiny images.
\newblock {\em Learning Multiple Layers of Features from Tiny Images}, 2009.

\bibitem{li2021sample}
Anran Li, Lan Zhang, Juntao Tan, Yaxuan Qin, Junhao Wang, and Xiang-Yang Li.
\newblock Sample-level data selection for federated learning.
\newblock In {\em IEEE INFOCOM 2021-IEEE Conference on Computer
  Communications}, pages 1--10. IEEE, 2021.

\bibitem{li2020federated}
Tian Li, Anit~Kumar Sahu, Ameet Talwalkar, and Virginia Smith.
\newblock Federated learning: Challenges, methods, and future directions.
\newblock {\em IEEE Signal Processing Magazine}, 37(3):50--60, 2020.

\bibitem{lim2020federated}
Wei Yang~Bryan Lim, Nguyen~Cong Luong, Dinh~Thai Hoang, Yutao Jiao, Ying-Chang
  Liang, Qiang Yang, Dusit Niyato, and Chunyan Miao.
\newblock Federated learning in mobile edge networks: A comprehensive survey.
\newblock {\em IEEE Communications Surveys \& Tutorials}, 22(3):2031--2063,
  2020.

\bibitem{lin2020ensemble}
Tao Lin, Lingjing Kong, Sebastian~U Stich, and Martin Jaggi.
\newblock Ensemble distillation for robust model fusion in federated learning.
\newblock {\em Advances in Neural Information Processing Systems}, 2020.

\bibitem{liu2022privacy}
Ken Liu, Shengyuan Hu, Steven~Z Wu, and Virginia Smith.
\newblock On privacy and personalization in cross-silo federated learning.
\newblock {\em Advances in neural information processing systems},
  35:5925--5940, 2022.

\bibitem{liu2019comparison}
Xiaoxuan Liu, Livia Faes, Aditya~U Kale, Siegfried~K Wagner, Dun~Jack Fu, Alice
  Bruynseels, Thushika Mahendiran, Gabriella Moraes, Mohith Shamdas, Christoph
  Kern, et~al.
\newblock A comparison of deep learning performance against health-care
  professionals in detecting diseases from medical imaging: a systematic review
  and meta-analysis.
\newblock {\em The lancet digital health}, 1(6):e271--e297, 2019.

\bibitem{liu2014efficient}
Yong Liu, Shali Jiang, and Shizhong Liao.
\newblock Efficient approximation of cross-validation for kernel methods using
  bouligand influence function.
\newblock In {\em ICML}, 2014.

\bibitem{liu2020learning}
Yuhan Liu, Ananda~Theertha Suresh, Felix Xinnan~X Yu, Sanjiv Kumar, and Michael
  Riley.
\newblock Learning discrete distributions: user vs item-level privacy.
\newblock {\em Advances in Neural Information Processing Systems},
  33:20965--20976, 2020.

\bibitem{mcmahan2017communication}
Brendan McMahan, Eider Moore, Daniel Ramage, Seth Hampson, and Blaise~Aguera
  y~Arcas.
\newblock Communication-efficient learning of deep networks from decentralized
  data.
\newblock In {\em Artificial intelligence and statistics}, pages 1273--1282.
  PMLR, 2017.

\bibitem{brendan2018learning}
H.~Brendan McMahan, Daniel Ramage, Kunal Talwar, and Li~Zhang.
\newblock Learning differentially private recurrent language models.
\newblock In {\em International Conference on Learning Representations}, 2018.

\bibitem{mohanty2016using}
Sharada~P Mohanty, David~P Hughes, and Marcel Salath{\'e}.
\newblock Using deep learning for image-based plant disease detection.
\newblock {\em Frontiers in plant science}, 7:1419, 2016.

\bibitem{oakden2020hidden}
Luke Oakden-Rayner, Jared Dunnmon, Gustavo Carneiro, and Christopher R{\'e}.
\newblock Hidden stratification causes clinically meaningful failures in
  machine learning for medical imaging.
\newblock In {\em Proceedings of the ACM conference on health, inference, and
  learning}, pages 151--159, 2020.

\bibitem{oala2023dmlr}
Luis Oala, Manil Maskey, Lilith Bat-Leah, Alicia Parrish, Nezihe~Merve
  G{\"u}rel, Tzu-Sheng Kuo, Yang Liu, Rotem Dror, Danilo Brajovic, Xiaozhe Yao,
  et~al.
\newblock Dmlr: Data-centric machine learning research--past, present and
  future.
\newblock {\em arXiv preprint arXiv:2311.13028}, 2023.

\bibitem{ogawa2013safe}
Kohei Ogawa, Yoshiki Suzuki, and Ichiro Takeuchi.
\newblock Safe screening of non-support vectors in pathwise svm computation.
\newblock In {\em ICML}, 2013.

\bibitem{rajpurkar2017chexnet}
Pranav Rajpurkar, Jeremy Irvin, Kaylie Zhu, Brandon Yang, Hershel Mehta, Tony
  Duan, Daisy Ding, Aarti Bagul, Curtis Langlotz, Katie Shpanskaya, et~al.
\newblock Chexnet: Radiologist-level pneumonia detection on chest x-rays with
  deep learning.
\newblock {\em arXiv preprint arXiv:1711.05225}, 2017.

\bibitem{so2020byzantine}
Jinhyun So, Ba{\c{s}}ak G{\"u}ler, and A~Salman Avestimehr.
\newblock Byzantine-resilient secure federated learning.
\newblock {\em IEEE Journal on Selected Areas in Communications}, 2020.

\bibitem{tang2017privacy}
Jun Tang, Aleksandra Korolova, Xiaolong Bai, Xueqiang Wang, and Xiaofeng Wang.
\newblock Privacy loss in apple's implementation of differential privacy on
  macos 10.12.
\newblock {\em arXiv preprint arXiv:1709.02753}, 2017.

\bibitem{triastcyn2020data}
Aleksei Triastcyn.
\newblock {\em Data-Aware Privacy-Preserving Machine Learning}.
\newblock PhD thesis, EPFL, Lausanne, 2020.

\bibitem{DBLP:conf/bigdataconf/TriastcynF19}
Aleksei Triastcyn and Boi Faltings.
\newblock Federated learning with bayesian differential privacy.
\newblock In {\em {IEEE} International Conference on Big Data (Big Data)}.
  {IEEE}, 2019.

\bibitem{tuor2021overcoming}
Tiffany Tuor, Shiqiang Wang, Bong~Jun Ko, Changchang Liu, and Kin~K Leung.
\newblock Overcoming noisy and irrelevant data in federated learning.
\newblock In {\em 25th International Conference on Pattern Recognition (ICPR)}.
  IEEE, 2021.

\bibitem{wang2021field}
Jianyu Wang, Zachary Charles, Zheng Xu, Gauri Joshi, H~Brendan McMahan, Maruan
  Al-Shedivat, Galen Andrew, Salman Avestimehr, Katharine Daly, Deepesh Data,
  et~al.
\newblock A field guide to federated optimization.
\newblock {\em arXiv preprint arXiv:2107.06917}, 2021.

\bibitem{wang2019beyond}
Zhibo Wang, Mengkai Song, Zhifei Zhang, Yang Song, Qian Wang, and Hairong Qi.
\newblock Beyond inferring class representatives: User-level privacy leakage
  from federated learning.
\newblock In {\em IEEE INFOCOM 2019-IEEE conference on computer
  communications}, pages 2512--2520. IEEE, 2019.

\bibitem{warner1965randomized}
Stanley~L Warner.
\newblock Randomized response: A survey technique for eliminating evasive
  answer bias.
\newblock {\em Journal of the American Statistical Association},
  60(309):63--69, 1965.

\bibitem{watson2022differentially}
Lauren Watson, Rayna Andreeva, Hao-Tsung Yang, and Rik Sarkar.
\newblock Differentially private shapley values for data evaluation.
\newblock {\em arXiv preprint arXiv:2206.00511}, 2022.

\bibitem{wu2020visual}
Bichen Wu, Chenfeng Xu, Xiaoliang Dai, Alvin Wan, Peizhao Zhang, Zhicheng Yan,
  Masayoshi Tomizuka, Joseph Gonzalez, Kurt Keutzer, and Peter Vajda.
\newblock Visual transformers: Token-based image representation and processing
  for computer vision, 2020.

\bibitem{xue2021toward}
Yihao Xue, Chaoyue Niu, Zhenzhe Zheng, Shaojie Tang, Chengfei Lyu, Fan Wu, and
  Guihai Chen.
\newblock Toward understanding the influence of individual clients in federated
  learning.
\newblock In {\em Proceedings of the AAAI Conference on Artificial
  Intelligence}, volume~35, pages 10560--10567, 2021.

\bibitem{yan2020evaluating}
Tom Yan, Christian Kroer, and Alexander Peysakhovich.
\newblock Evaluating and rewarding teamwork using cooperative game
  abstractions.
\newblock {\em Advances in Neural Information Processing Systems},
  33:6925--6935, 2020.

\bibitem{yang2019federated}
Qiang Yang, Yang Liu, Tianjian Chen, and Yongxin Tong.
\newblock Federated machine learning: Concept and applications.
\newblock {\em ACM Transactions on Intelligent Systems and Technology (TIST)},
  10(2):1--19, 2019.

\bibitem{pmlr-v80-yin18a}
Dong Yin, Yudong Chen, Ramchandran Kannan, and Peter Bartlett.
\newblock {B}yzantine-robust distributed learning: Towards optimal statistical
  rates.
\newblock In Jennifer Dy and Andreas Krause, editors, {\em Proceedings of the
  35th International Conference on Machine Learning}, Proceedings of Machine
  Learning Research. PMLR, 2018.

\bibitem{opacus}
Ashkan Yousefpour, Igor Shilov, Alexandre Sablayrolles, Davide Testuggine,
  Karthik Prasad, Mani Malek, John Nguyen, Sayan Ghosh, Akash Bharadwaj,
  Jessica Zhao, Graham Cormode, and Ilya Mironov.
\newblock Opacus: {U}ser-friendly differential privacy library in {PyTorch}.
\newblock {\em arXiv preprint arXiv:2109.12298}, 2021.

\bibitem{DBLP:journals/corr/abs-2109-12298}
Ashkan Yousefpour, Igor Shilov, Alexandre Sablayrolles, Davide Testuggine,
  Karthik Prasad, Mani Malek, John Nguyen, Sayan Ghosh, Akash Bharadwaj,
  Jessica Zhao, Graham Cormode, and Ilya Mironov.
\newblock Opacus: User-friendly differential privacy library in pytorch.
\newblock {\em CoRR}, abs/2109.12298, 2021.

\bibitem{yu2022tct}
Yaodong Yu, Alexander Wei, Sai~Praneeth Karimireddy, Yi~Ma, and Michael Jordan.
\newblock {TCT}: Convexifying federated learning using bootstrapped neural
  tangent kernels.
\newblock In Alice~H. Oh, Alekh Agarwal, Danielle Belgrave, and Kyunghyun Cho,
  editors, {\em Advances in Neural Information Processing Systems}, 2022.

\end{thebibliography}
\clearpage

\section{Appendix}

\section*{Contents}

This appendix covers further details of our work
. Specifically:

\begin{enumerate}
    \item In Section~\ref{sec:Ap Setting} we describe the methodology of our benchmarks. 
    \item In Section~\ref{sec:Ap Implementation} we discuss the datasets used, the selected hyper-parameters, and other implementation details. 
    \item In Section~\ref{sec:Ap Social} we describe the potential positive and negative societal impact of our work.
    \item In Section~\ref{sec:Ap Limitations} we briefly discuss the limitations of our work.
    \item In Sections~\ref{sec: numerical results} we provide additional discussion of our simulation results.
\end{enumerate}

\section{Setting}  \label{sec:Ap Setting}

We consider a classification problem from some input space $\mathcal{X}$ (e.g., features, images, etc.) to an output space $\mathcal{Y}$ (e.g., labels). In a Federated Learning setting, there is a center $C$ that wants to learn a model $M(\theta)$ parameterized by $\theta \in \Theta$, with a non-negative loss function $L(z, \theta)$ on a sample $z=(\bar{x},y) \in \mathcal{X} \times \mathcal{Y}$. Let $R(Z, \theta) = \frac{1}{n}\sum_{i=1}^n L(z_i, \theta)$ denote the empirical risk, given a set of data $Z = \{z_i\}_{i=1}^n$. We assume that the empirical risk is differentiable in $\theta$.The training data are supplied by a set of data holders.

\begin{figure*}[h]
	\centering
	\begin{subfigure}{.33\linewidth}
	  \centering
	  \includegraphics[width=\linewidth, clip, trim={3em 0.5em 3em 2.5em}]{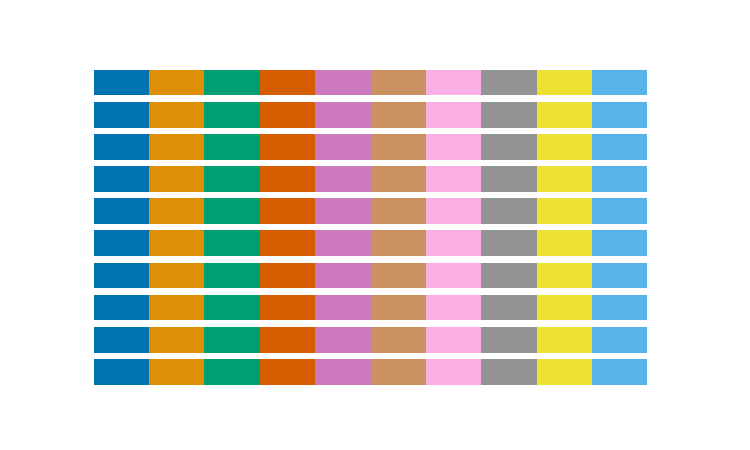}
 	  \caption{$\bm{\alpha} \rightarrow \infty$}
	\end{subfigure}%
	\begin{subfigure}{.33\linewidth}
	  \centering
	  \includegraphics[width=\linewidth, clip, trim={3em 0.5em 3em 2.5em}]{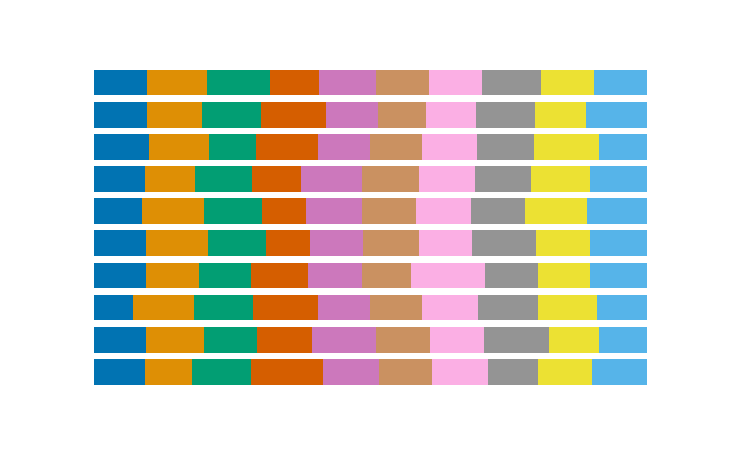}
	  \caption{$\bm{\alpha} \rightarrow 100$}
	\end{subfigure}%
	\begin{subfigure}{.33\linewidth}
	  \centering
	  \includegraphics[width=\linewidth, clip, trim={3em 0.5em 3em 2.5em}]{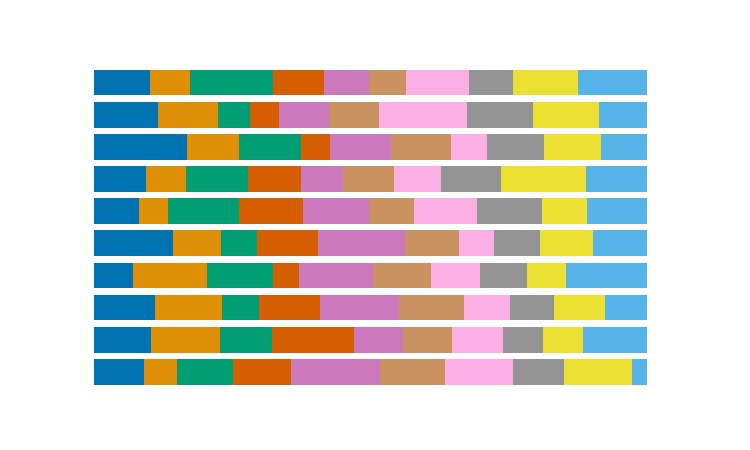}
	  \caption{$\bm{\alpha} \rightarrow 10$}
	\end{subfigure}

	\begin{subfigure}{.33\linewidth}
	  \centering
	  \includegraphics[width=\linewidth, clip, trim={3em 0.5em 3em 2.5em}]{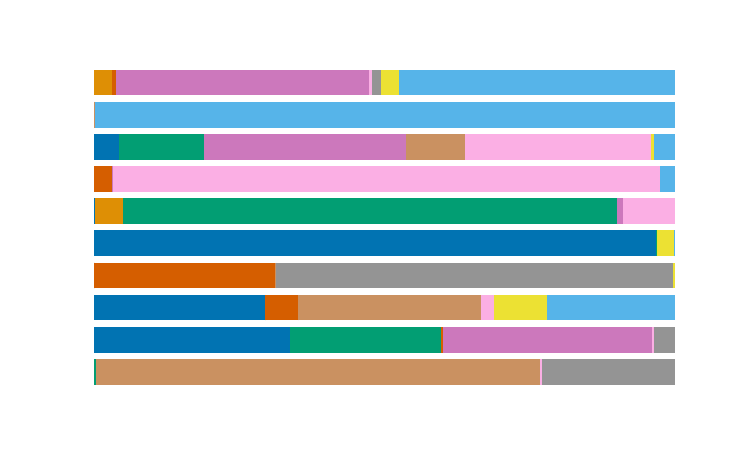}
	  \caption{$\bm{\alpha} \rightarrow 0.1$}
	\end{subfigure}%
	\begin{subfigure}{.33\linewidth}
	  \centering
	  \includegraphics[width=\linewidth, clip, trim={3em 0.5em 3em 2.5em}]{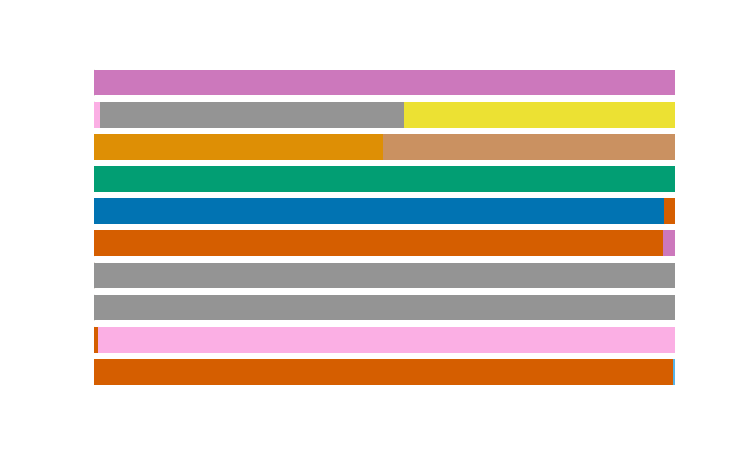}
	  \caption{$\bm{\alpha} \rightarrow 0.01$}
	\end{subfigure}%
	\begin{subfigure}{.33\linewidth}
	  \centering
	  \includegraphics[width=\linewidth, clip, trim={3em 0.5em 3em 2.5em}]{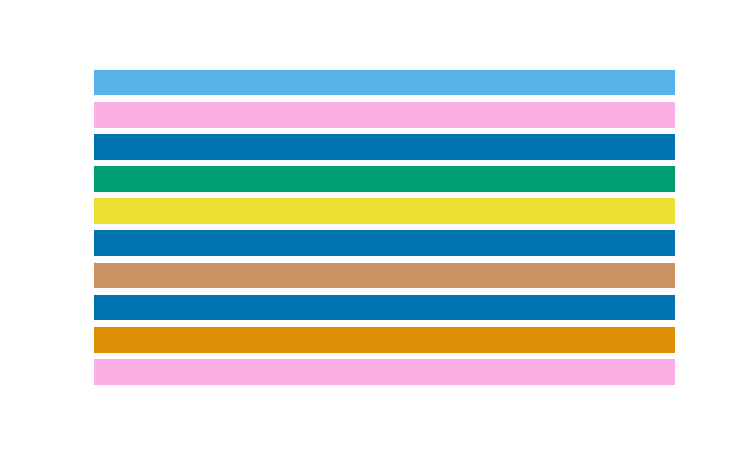}
	  \caption{$\bm{\alpha} \rightarrow 0$}
	\end{subfigure}
 
	\caption{Dirichlet distribution visualisation for 10 classes, parametrized by $\bm{\alpha}$. $\bm{\alpha}$ controls the concentration of different classes. Each row represents a participant, each color a different class, and each colored segment the amount of data the participant has from each class. For $\bm{\alpha} \rightarrow \infty$, each participant has the same amount of data from each class (IID distribution). For $\bm{\alpha} \rightarrow 0$, each participant only holds data from one class. In this work, we use $\bm{\alpha} \rightarrow 0.1$ for a non-IID distribution.}
	\label{Dirichlet}
\end{figure*}

\subsection{Non-IID Setting}  \label{sec: Appendix-Non-IID Setting}

The main hurdle for Federated Learning is that not all data is IID. Heterogeneous data distributions are all but uncommon in the real world. To simulate a Non-IID distribution, we used Dirichlet distribution to split the training dataset as in related literature~\cite{hsu2019measuring,lin2020ensemble,hoech2022,yu2022tct}. This distribution is parameterized by $\alpha$, which controls the concentration of different classes, as visualized in Figure~\ref{Dirichlet}. This work uses $\alpha \rightarrow 0.1$ for a non-IID distribution, as in related literature (e.g., \cite{yu2022tct}).

\subsection{Exact Influence}  \label{sec: Exact-Influence}

In simple terms, influence measures the marginal contribution of a data point on a model's accuracy. A positive influence value indicates that a data point improves model accuracy, and vice-versa. More specifically, let $Z = \{z_i\}_{i=1}^n$,  $Z_{+j} = Z \cup z_j$ where $z_j \not\in Z$, and let
\begin{equation*}
 \hat{R} = \min_{\theta} R(Z, \theta) \text{\quad and\quad} \hat{R}_{+j} = \min_{\theta} R(Z_{+j}, \theta)
\end{equation*}

\noindent
where $\hat{R}$ and $\hat{R}_{+j}$ denote the minimum empirical risk of their respective set of data. The \emph{influence} of datapoint $z_j$ on $Z$ is defined as:
\begin{equation} \label{eq:exact-influence}
    \mathcal{I}(z_j,Z) \triangleq \hat{R} - \hat{R}_{+j}
\end{equation}

Despite being highly informative, influence functions have not achieved widespread use in Federated Learning (or Machine Learning in general). This is mainly due to the computational cost. Equation \ref{eq:exact-influence} requires complete retraining of the model, which is time-consuming, and very costly; especially for state-of-the-art, large ML models. Moreover, specifically in our setting, we do not have direct access to the training data. In the following section, we will introduce a practical approximation of the influence, applicable in Federated Learning scenarios.

\subsection{Influence Approximation}  \label{sec: Influence Approximation}

The first-order Taylor approximation of influence, adopted by~\cite{pmlrv70koh17a} (based on~\cite{cook1982residuals}), to understand the effects of training points on the predictions of a \emph{centralized} ML model. To the best of our knowledge, this is the current state-of-the-art approach to utilizing the influence function in ML. Thus, it is worth taking the time to understand the challenges that arise if we adopt this approximation in the Federated Learning setting. 

Let $\hat{\theta} = \arg\min_{\theta} R(Z, \theta)$ denote the empirical risk minimizer. The approximate influence of a training point $z_{j}$ on the validation point $z_{val}$ can be computed without having to re-train the model, according to the following equation:
\begin{equation} \label{eq: influence approximation Koh}
    \mathcal{I}_{appr}(z_{j}, z_{val}) \triangleq - \nabla_\theta L(z_{val}, \hat{\theta}) H^{-1}_{\hat{\theta}} \nabla_\theta L(z_j, \hat{\theta})
\end{equation}

\noindent
where $H^{-1}_{\hat{\theta}}$ is the inverse Hessian computed on all the model's training data. The advantage of Equation \ref{eq: influence approximation Koh} is that we can answer counterfactuals on the effects of up/down-scaling a training point, without having to re-train the model. One can potentially average over the validation points of a participant, and/or across the training points in a batch of a contributor, to get the total influence.

\subsection{Challenges}

\begin{figure*}
\centering
\begin{subfigure}{.45\textwidth}
  \centering
  \includegraphics[width=\linewidth]{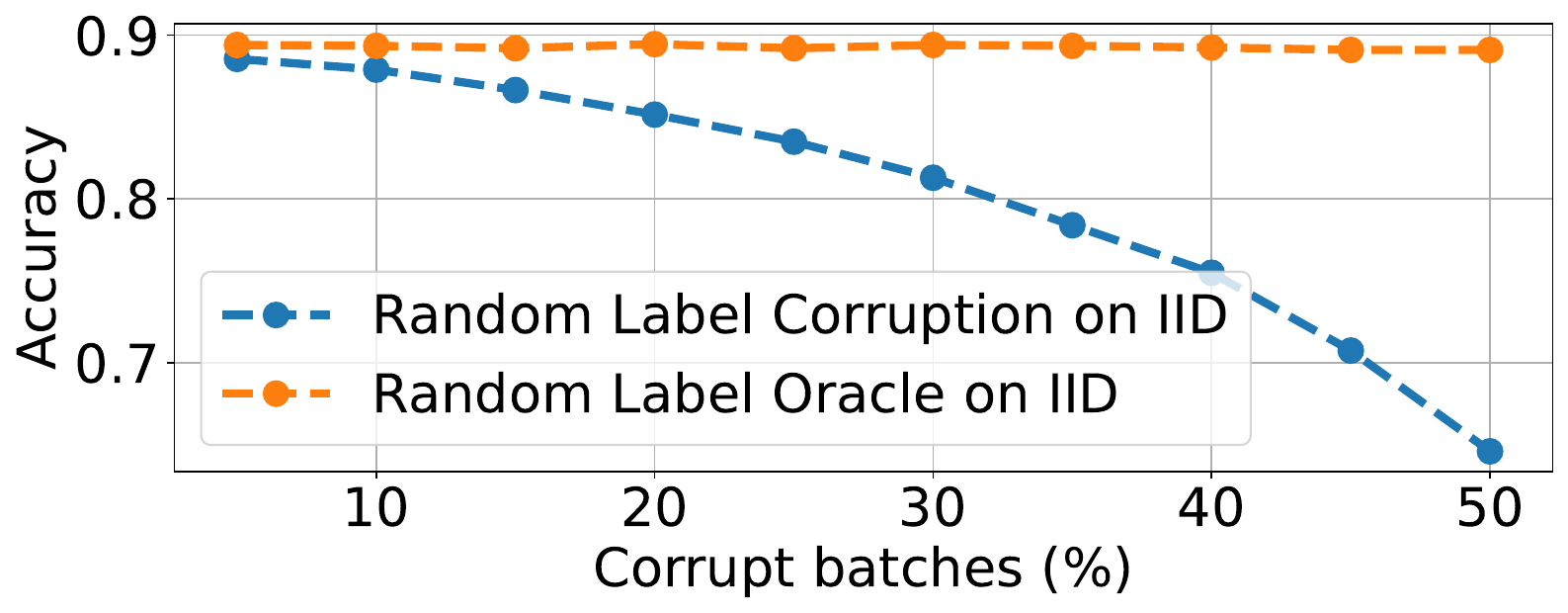}
  \caption{Random Label IID}
    \label{fig:random_label_percent}
\end{subfigure}%
\begin{subfigure}{.45\textwidth}
  \centering
  \includegraphics[width=\linewidth]{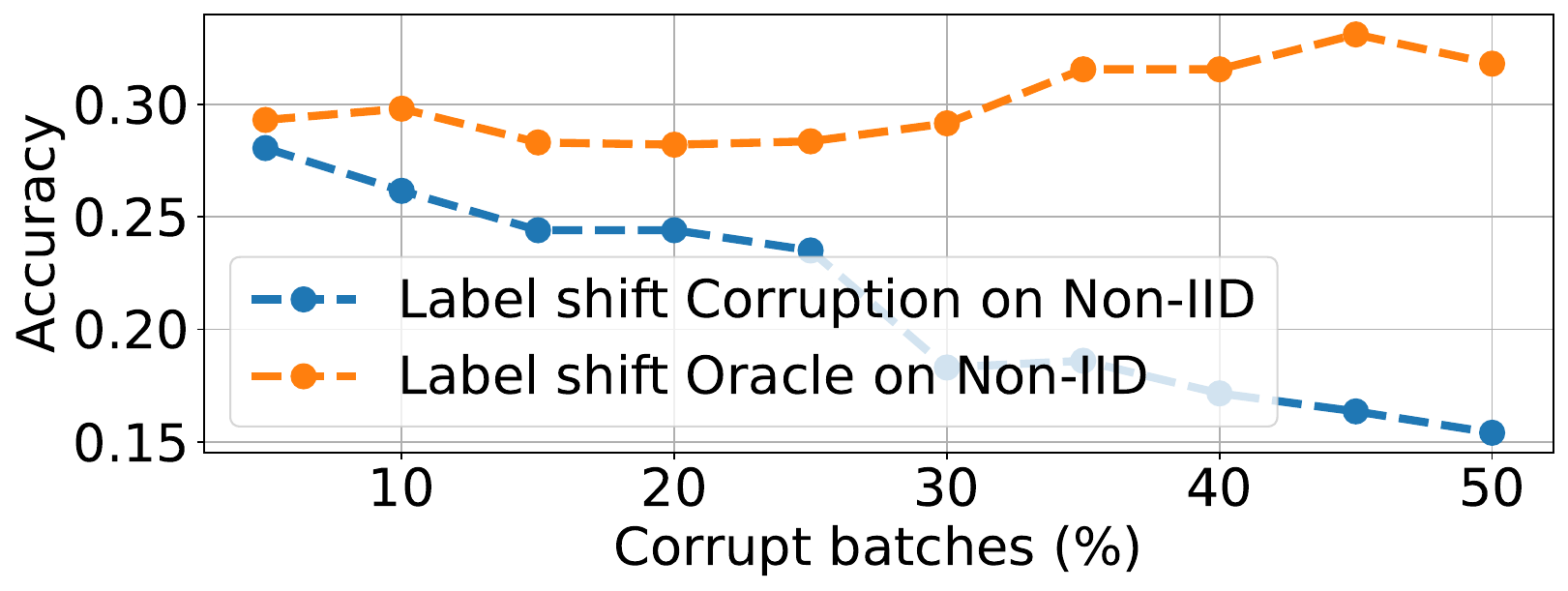}
  \caption{Label Shift Non-IID ($\alpha = 0.01$)}
    \label{fig:label_shift_percent}
\end{subfigure}

\caption{Model accuracy relative to different mislabel rates (5\% - 50\%). These models have been trained over 25 communication rounds and 100 participants. We compare a centralized model with no filtering of mislabeled data (\textcolor{figure_blue}{blue}) to an FL model under perfect (oracle) filtering (\textcolor{figure_orange}{orange}). Note that the lower accuracy on the Non-IID setting is due to the fact that we are considering the most extreme non-IID case. This is where the majority of the participants have access to at most 1 class.}
\label{fig:motivation_corruption}
\end{figure*}

Consider Figure \ref{fig:motivation_corruption} as a motivating example. In this scenario, we have participants with corrupted data. Even a very robust model (ViT) loses performance when corruption is involved. This can also be observed in the work of~\cite{li2021sample}. Filtering those corrupted participants (orange line) restores the model's performance.

While Equation \ref{eq: influence approximation Koh} can be an effective tool in understanding centralized machine learning systems, it is \emph{ill-matched} for Federated Learning models, for several key reasons. 

To begin with, evaluating Equation \ref{eq: influence approximation Koh} requires \emph{forming and inverting} the Hessian of the empirical risk. With $n$ training points and $\theta \in \mathbb{R}^m$, this requires $O(nm^2 + m^3)$ operations~\cite{pmlrv70koh17a}, which is \emph{impractical} for modern-day deep neural networks with millions of parameters. To overcome these challenges, \cite{pmlrv70koh17a} used implicit Hessian-vector products (HVPs) to more efficiently approximate $\nabla_\theta L(z_{val}, \hat{\theta}) H^{-1}_{\hat{\theta}}$, which typically requires $O(p)$ \cite{pmlrv70koh17a}. While this is a somewhat more efficient computation, it is \emph{communication-intensive}, as it requires \emph{transferring all of the (either training or validation) data} at each FL round. Most importantly, it \emph{can not provide any privacy} to the users' data, an important, inherent requirement/constraint in FL.

Finally, to compute Equation \ref{eq: influence approximation Koh}, the loss function has to be strictly convex and twice differentiable (which is not always the case in modern ML applications). A proposed solution is to swap out non-differentiable components for smoothed approximations~\cite{pmlrv70koh17a}, but there is no quality guarantee of the influence calculated in this way.

\section{Implementation details}\label{sec:Ap Implementation}

This section describes the base model used in our simulations and all hyper-parameters. Specifically, we used a Visual Image Transformer (VIT)~\cite{deng2009imagenet,wu2020visual}. The basis of our model represents a model pre-trained on  ImageNet-21k at 224x224 resolution and then fine-tuned on ImageNet 2012 at 224x224 resolution. All hyper-parameters added or changed from the default VIT hyper-parameters are listed in Table~\ref{tab: hyper-params} with their default values.
The following hyper-parameters have been added to support our evaluation technique:
\begin{itemize}
    \item \textbf{Random Factor}: this coefficient represents the amount of corrupted data inside a corrupted batch.
    \item \textbf{Final Evaluation Size}: an a priori separated batch of test data to evaluate model performance.
    \item \textbf{Parameters to Change}: number of parameters (and biases) in the last layer of the model.
\end{itemize}

For the HAR dataset we have used a simple two-layer fully connected neural network. This network has not been pretrained like the previous one. With this network we chose to omit regularization and the detailed hyper-parameters are listed in Table~\ref{tab: hyper-params}. \cite{DBLP:journals/corr/abs-2109-12298}

Regarding reproducibility, we ran the provided (in the supplementary material) code for each dataset with seeds from the range of $0-7$.

\begin{table}[]
\centering
\begin{tabular}{@{}llll@{}}
\toprule
                      & CIFAR10 & CIFAR100 & HAR\\ \midrule
No. of Participants   & 100     & 100      & 500\\
Batch Size            & 100     & 250      & 2000\\
Validation Size       & 50      & 50       & 500\\
Random Factor         & 0.9     & 0.9      & 1.0\\
Warm-up Size          & 600     & 4000     & 7352\\
Final Evaluation Size & 2000    & 2000     & 2947\\
Load Best Model       & False   & False    & False\\
Parameters to Change  & 7690    & 76900    & 36358\\
Learning Rate         & 0.001   & 0.001    & 0.001\\
Train Epochs          & 3       & 3        &  20\\
Weight Decay          & 0.01    & 0.01     & None\\ \bottomrule
\end{tabular}
\caption{Table of hyper-parameters.}
\label{tab: hyper-params}
\end{table}

\subsection{Termination Condition}
Different termination conditions have been used for our proposed solution and to retrain the exact influence. Our solution has only one termination condition, that is the number of local epochs $k$.

\subsection{Computational Resources} \label{Ap: Computational Resources}
All simulations were run on two different systems:
\begin{enumerate}
    \item Intel Xeon E5-2680 – 12 cores, 24 threads, 2.5 GHz – with 256 GB of RAM, Titan X GPU (Pascal)
    \item EPFL RCP Cluster with a100 and h100 GPUs.
\end{enumerate}

\section{Societal Impact} \label{sec:Ap Social}

Privacy advocacy movements have, in recent years, raised their voices about the potential abuse of these systems. Additionally, legal institutions have also recognized the importance of privacy, and have passed regulations in accordance, for example, the General Data Protection Regulation (GDPR). Our work provides practical privacy guarantees to protect all parties, with minimal compromise on performance. Furthermore, we allow data holders and collectors to be paid for their contribution in a joint model, instead of simply taking the data. Such incentives could potentially help speed up the growth of underdeveloped countries, and provide more high-quality data to researchers (as an example application, consider paying low-income farmers for gathering data in crop disease prevention~\cite{mohanty2016using}).

\section{Limitations} \label{sec:Ap Limitations}

The main limitation of our approach is that if the optimizer does not produce a good enough gradient, we cannot get a good approximation of the direction the model is headed for. The result of this is a lower score, and therefore a potentially inaccurate prediction.

\begin{figure*}[t!]
\centering
\begin{subfigure}{.5\textwidth}
  \centering
  \includegraphics[width=\linewidth]{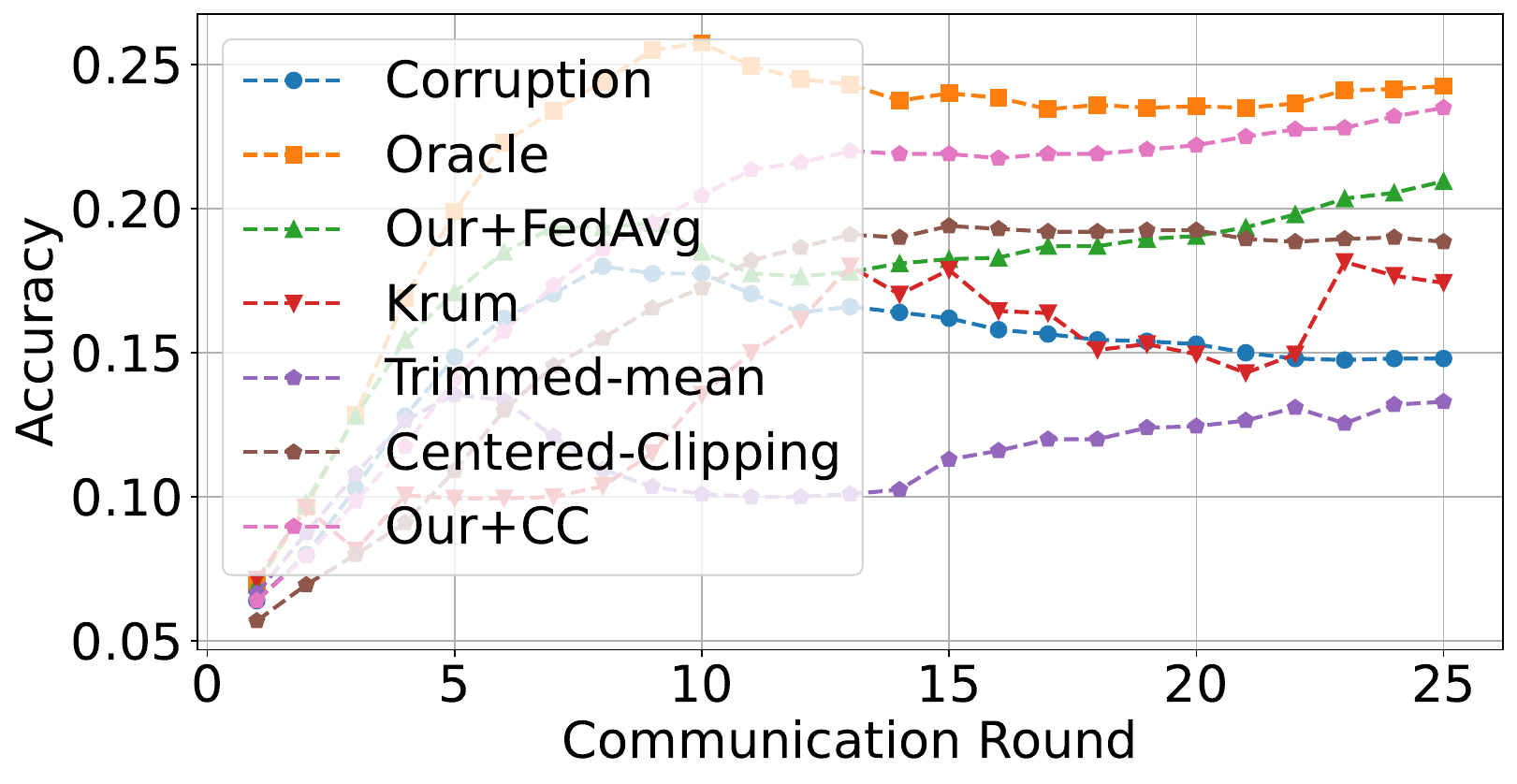}
  \caption{Label shift Non-IID ($\alpha = 0.01$)}
\end{subfigure}%
\begin{subfigure}{.5\textwidth}
  \centering
  \includegraphics[width=\linewidth]{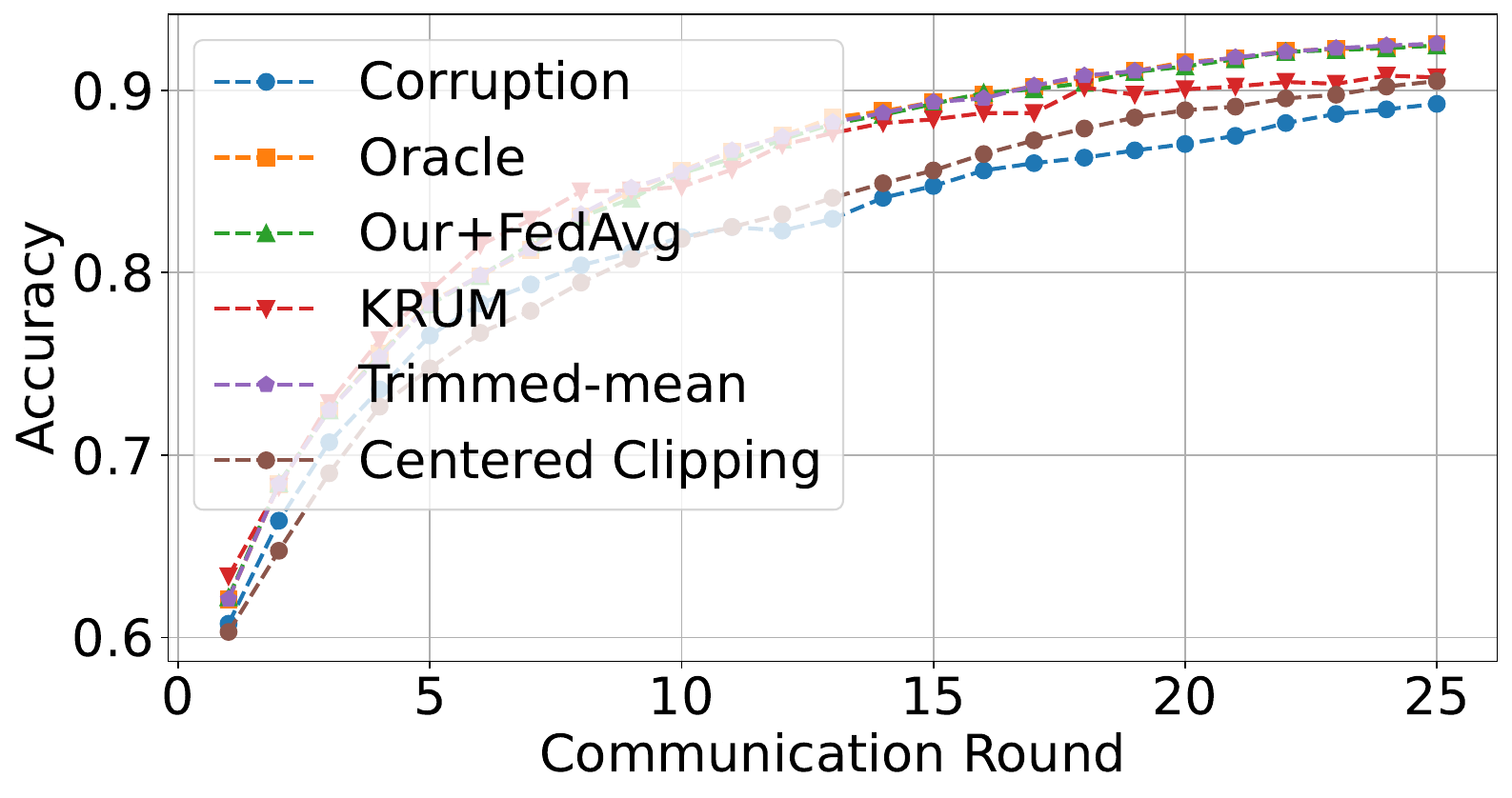}
  \caption{Random Label IID}
\end{subfigure}
\caption{Model accuracy over 25 communication rounds with a 30\% mislabel rate on CIFAR-10. We compare a centralized model with no filtering (\textcolor{figure_blue}{blue}) to an FL model under perfect (oracle) filtering (\textcolor{figure_orange}{orange}),  KRUM (\textcolor{figure_red}{red}), Trimmed-mean (\textcolor{figure_purple}{purple}), and our approach (\textcolor{figure_green}{green}). Note that the jagged line for KRUM is because only a single gradient is selected instead of performing FedAvg.}
\label{fig:Ap performance}
\end{figure*}

Another potential limitation is the filtering of ``good'' data. These data may be correctly labeled, but including it does not essentially provide any benefit to the model, as can be shown by the accuracy scores in Figure~\ref{fig:Ap performance}. While this allows us to train models of equal performance with a fraction of the data, some participants may be filtered out, even though they contribute accurate data. This might deter users from participating in the future.

\begin{table}[!]
\resizebox{1\linewidth}{!}{
\centering
\begin{tabular}{@{}l|ccccc|@{}}
\cmidrule(l){2-6}
                               & \multicolumn{5}{c|}{Corruption}                                                       \\
                               & 0\%             & 10\%            & 20\%            & 30\%           & 40\%           \\ \midrule
\multicolumn{1}{|r|}{Accuracy} & 85.2  +- 2.1 \% & 86.4 +- 1.2 \%  & 87.2 +-  1.3 \% & 87.4 +- 1.1 \% & 89.0 +- 1.2 \% \\
\multicolumn{1}{|r|}{Recall}   & -               & 97.0 +-  0.5 \% & 96.4 +- 0.7\%   & 98.1 +- 0.6\%  & 97.5 +- 0.7 \% \\ \bottomrule
\end{tabular}
}
\caption{Filtering performance on varying percentages of corrupt participants on the Human Activity Recognition dataset for \textit{highly non-IID}. Following the same strict \textit{worst-case differential privacy} guarantees ($\varepsilon \leq 1, \delta = 10^{-5}$).}
\label{tab:corruption_percent}
\end{table}

\begin{figure*}[t]
\centering
\begin{subfigure}{.33\textwidth}
  \centering
  \includegraphics[width=1\linewidth, height=9em, clip]{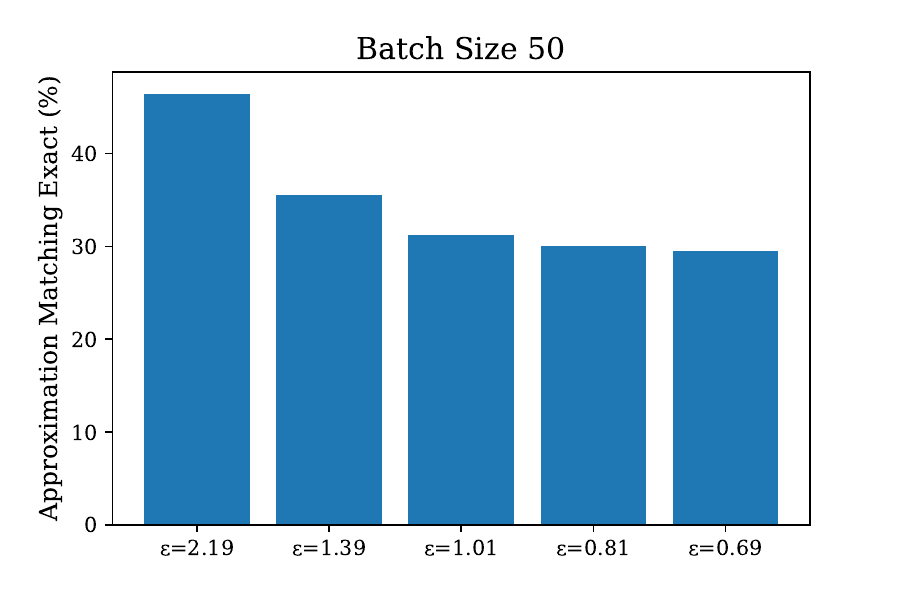}
  \caption{ }
  \label{fig:ex_sub1}
\end{subfigure}%
\begin{subfigure}{.33\textwidth}
  \centering
  \includegraphics[width=1\linewidth, height=9em, clip]{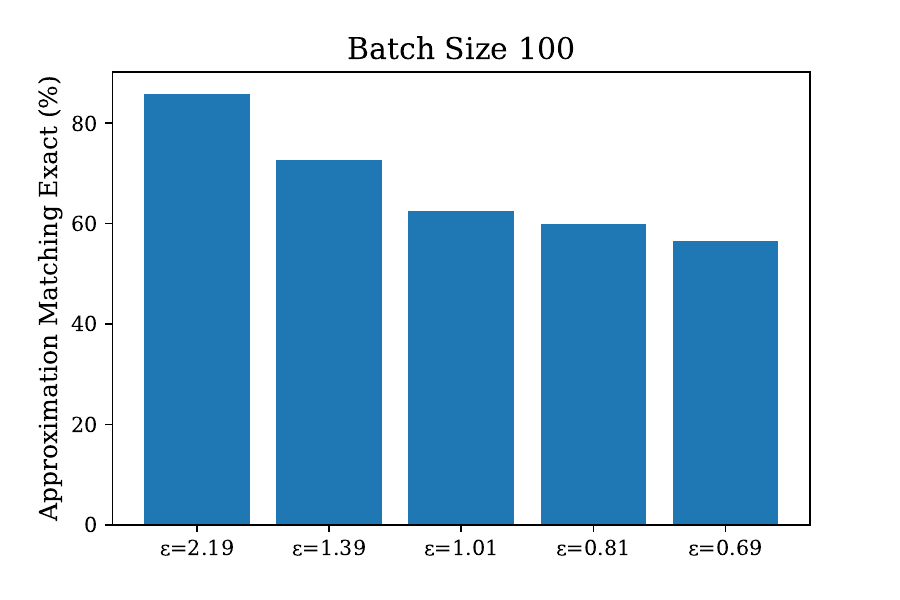}
  \caption{ }
  \label{fig:ex_sub2}
\end{subfigure}
\begin{subfigure}{.33\textwidth}
  \centering
  \includegraphics[width=1\linewidth, height=9em, clip]{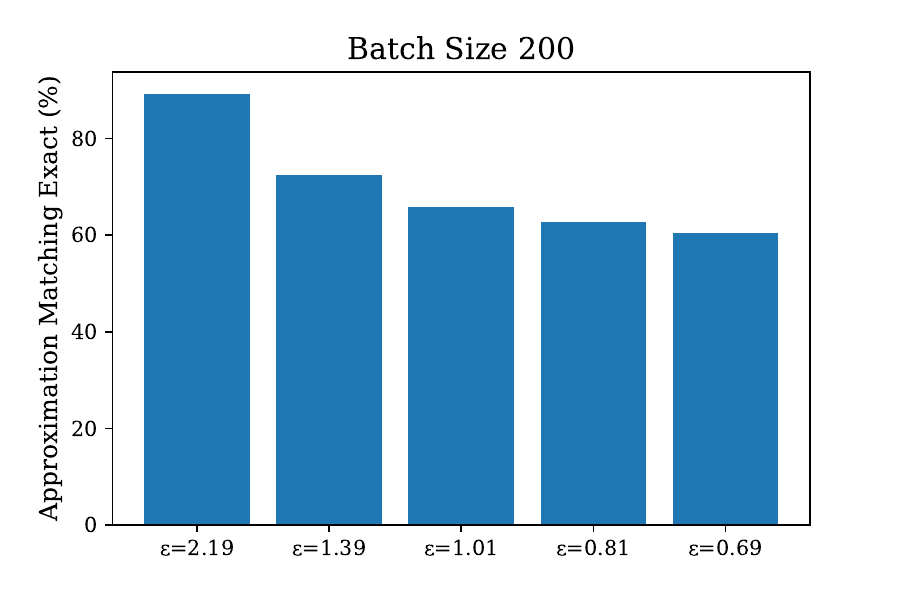}
  \caption{ }
  \label{fig:ex_sub3}
\end{subfigure}

\caption{Effect of batch size to the percentage of the times the sign of the proposed Lazy Influence Approximation (LIA) matches the sign of the exact influence, for varying differential privacy guarantees ($\varepsilon$ in the $x$-axis). Comparing \ref{fig:ex_sub1} to the other two figures, we can observe that there is a certain threshold of data that needs to be passed for \textit{lazy influence} to be effective. After this threshold has been reached, adding more data only gives marginal improvement as can be seen by comparing \ref{fig:ex_sub2} and \ref{fig:ex_sub3}.}
\label{fig:exact_infl_match}
\end{figure*}

\section{Numerical Results} \label{sec: numerical results}

\begin{table}[!t]
\resizebox{1\linewidth}{!}{%
\begin{tabular}{@{}cl|lll|@{}}
\cmidrule(l){3-5}
\multicolumn{1}{l}{}                             &              & \multicolumn{3}{c|}{Filtration Metrics}                                                        \\ \cmidrule(l){2-5} 
\multicolumn{1}{l|}{}                            & Distribution & \multicolumn{1}{l|}{Recall}           & \multicolumn{1}{l|}{Precision}       & Accuracy        \\ \midrule
\multicolumn{1}{|c|}{\multirow{2}{*}{CIFAR 10}}  & IID          & \multicolumn{1}{l|}{97.08 ± 3.51 \%}  & \multicolumn{1}{l|}{91.91 ± 7.15 \%} & 96.38 ± 2.83 \% \\
\multicolumn{1}{|c|}{}                           & Non-IID      & \multicolumn{1}{l|}{93.75 ± 5.12 \%}  & \multicolumn{1}{l|}{69.02 ± 6.28 \%} & 85.00 ± 3.28 \% \\ \midrule
\multicolumn{1}{|c|}{\multirow{2}{*}{CIFAR 100}} & IID          & \multicolumn{1}{l|}{99.17 ± 2.20 \%}  & \multicolumn{1}{l|}{97.96 ± 2.30 \%} & 99.12 ± 1.27 \% \\
\multicolumn{1}{|c|}{}                           & Non-IID      & \multicolumn{1}{l|}{92.50 ±  5.71 \%} & \multicolumn{1}{l|}{55.41 ± 3.94 \%} & 75.12 ± 3.76 \% \\ \midrule
\multicolumn{1}{|c|}{\multirow{2}{*}{HAR}} & IID          & \multicolumn{1}{l|}{100.00 ± 0.0\%}  & \multicolumn{1}{l|}{100.00 ± 0.0\%} & 100.00 ± 0.0\% \\
\multicolumn{1}{|c|}{}                           & Non-IID      & \multicolumn{1}{l|}{95.77 ± 2.1\%} & \multicolumn{1}{l|}{71.71 ± 1.3\%} & 87.40 ± 1.1\%\\ \bottomrule
\end{tabular}
}
\caption{Filtering performance on various datasets, including \emph{real-data} on Human Activity Recognition, for IID and \emph{highly non-IID} setting ($\alpha \rightarrow 0.1$, i.e., 3 classes per participant for HAR). 100 participants, 30\% mislabeling rate. \emph{Strict worst-case differential privacy} guarantees ($\varepsilon \leq 1, \delta = 10^{-5}$). }
\label{tab: Filtration metrics}
\end{table}

We provide detailed results that include both the means and standard deviations. The metrics can be found in Table~\ref{tab: Filtration metrics}.
The following subsections provide a more comprehensive analysis of the results, summarized in the main text due to space limitations.

\begin{figure*}[t]
	\centering
 	\begin{subfigure}{0.49\linewidth}
	  \centering
	  \includegraphics[width=\linewidth, clip, trim={0em 1em 0em 1em}]{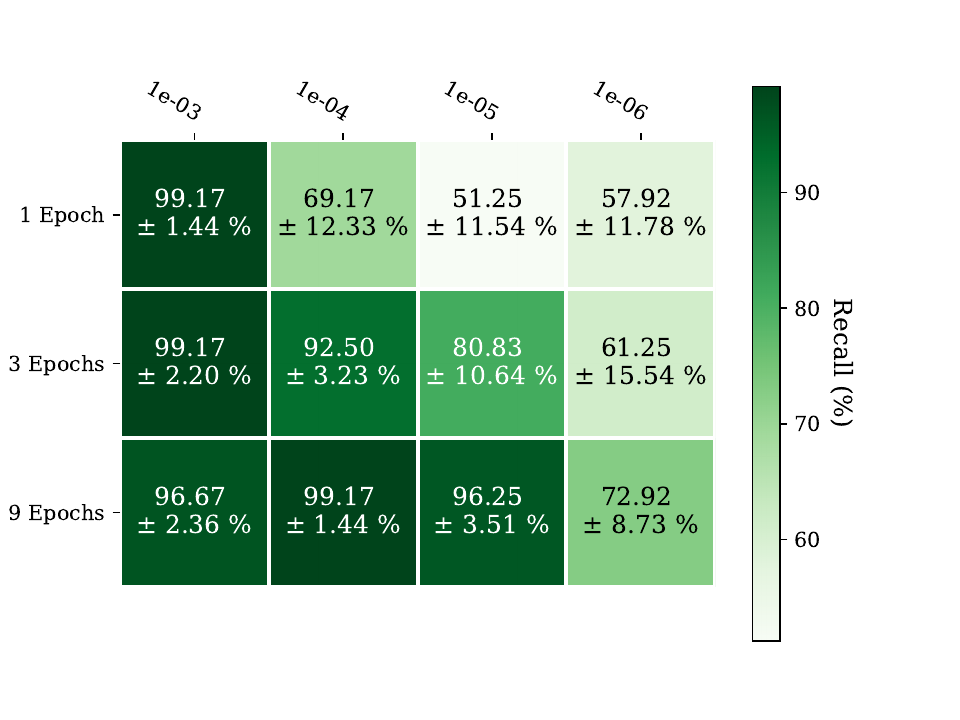}
	  \caption{Recall}
	\end{subfigure}
	\begin{subfigure}{0.49\linewidth}
	  \centering
	  \includegraphics[width=\linewidth, clip, trim={0em 1em 0em 1em}]{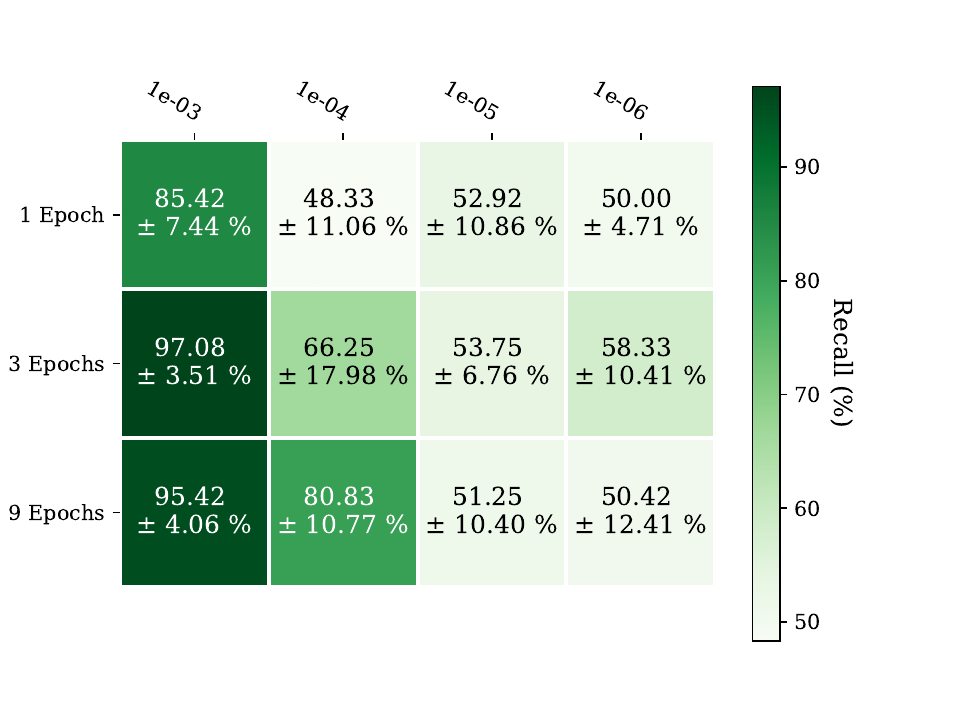}
	  \caption{Recall}
	\end{subfigure}
	\begin{subfigure}{0.49\linewidth}
	  \centering
	  \includegraphics[width=\linewidth, clip, trim={0em 1em 0em 0em}]{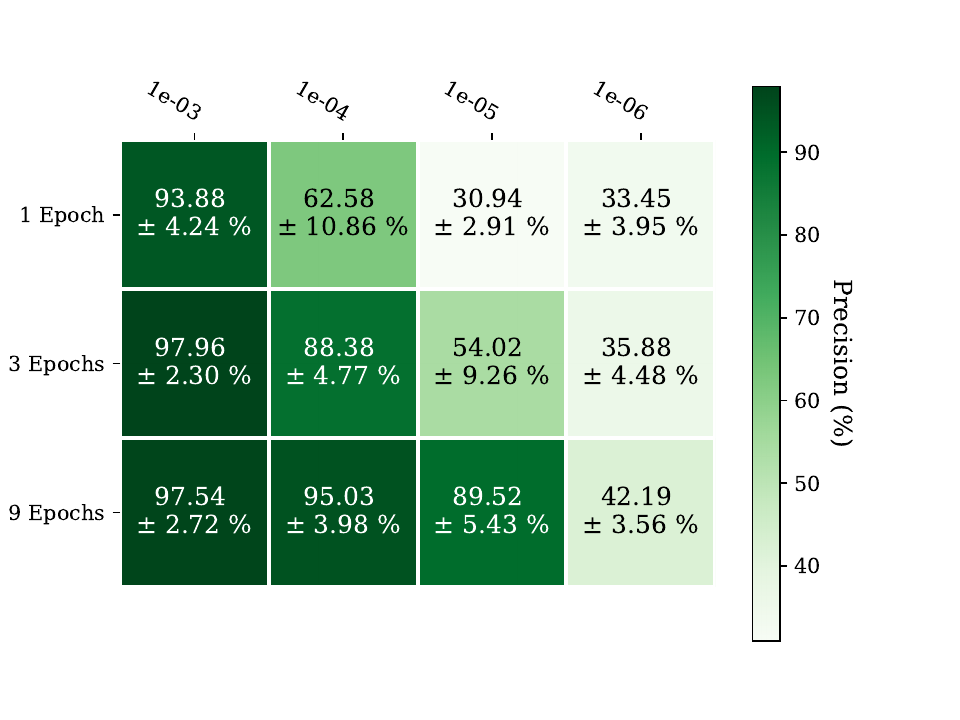}
	  \caption{Precision}
	\end{subfigure}
	\begin{subfigure}{0.49\linewidth}
	  \centering
	  \includegraphics[width=\linewidth, clip, trim={0em 1em 0em 0em}]{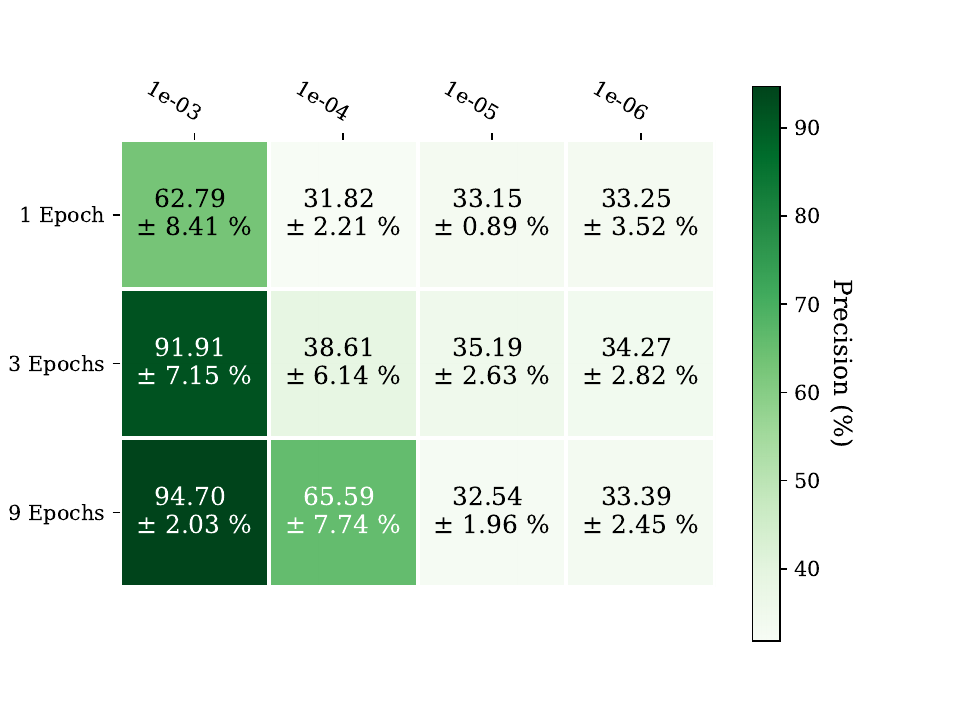}
	  \caption{Precision}
	\end{subfigure}
 	\begin{subfigure}{0.49\linewidth}
	  \centering
	  \includegraphics[width=\linewidth, clip, trim={0em 1em 0em 0em}]{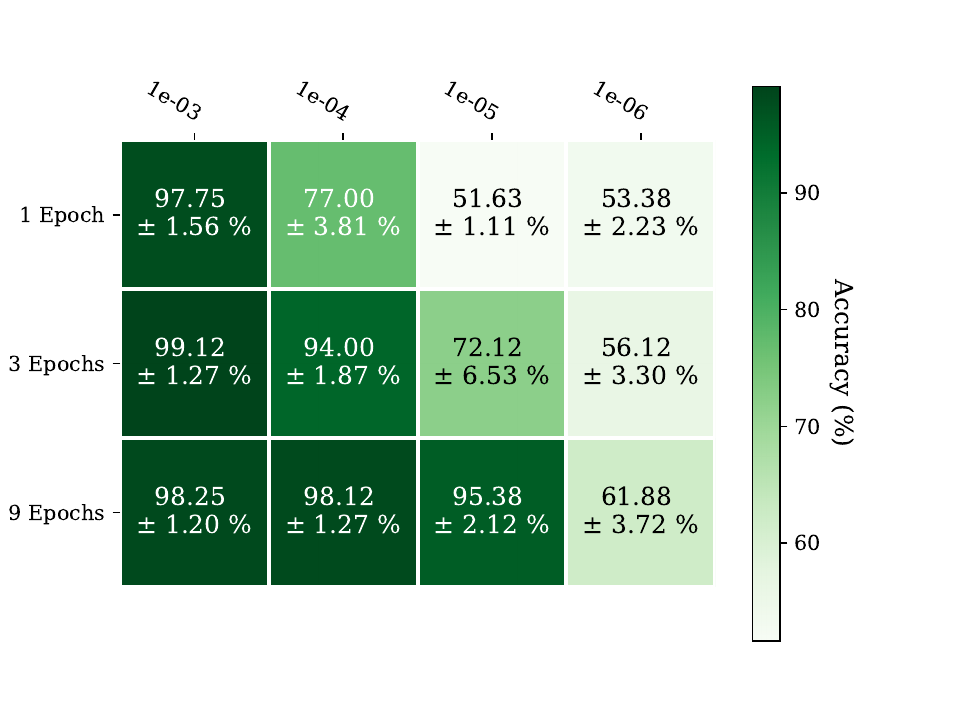}
	  \caption{Accuracy}
	\end{subfigure} 
	\begin{subfigure}{0.49\linewidth}
	  \centering
	  \includegraphics[width=\linewidth, clip, trim={0em 1em 0em 0em}]{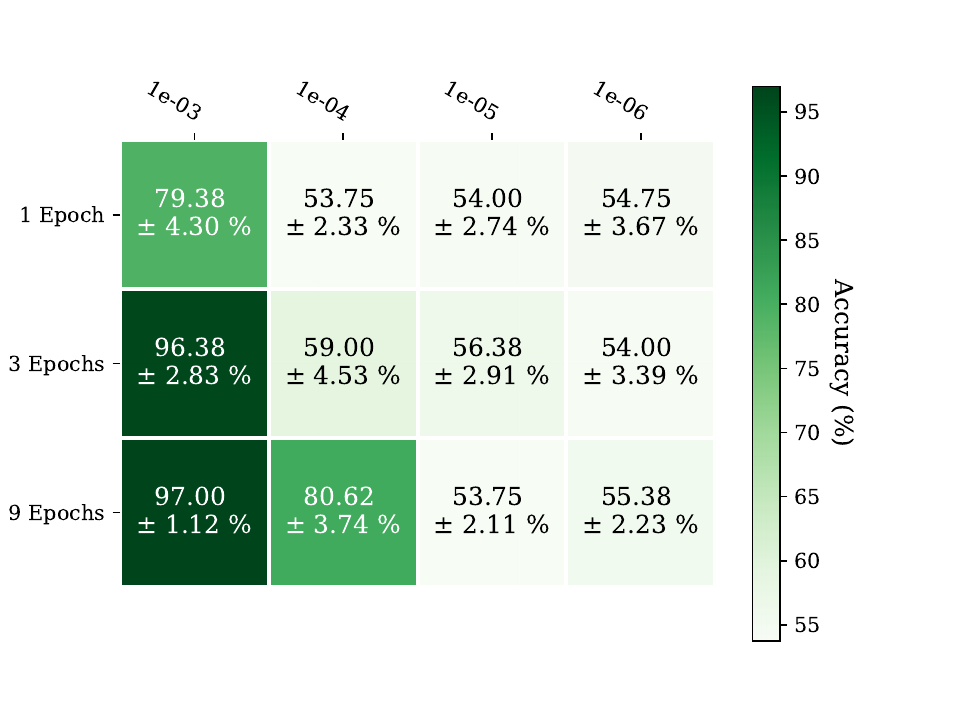}
	  \caption{Accuracy}
	\end{subfigure} 
	\caption{Recall (top), Precision (middle), and Accuracy(Bottom) on CIFAR 100 (left) and CIFAR 10 (right), IID, for different parameter pairs of learning rate and epoch count.}
\label{fig:iid c10}
\end{figure*}

\begin{figure*}[t]
	\centering
	\begin{subfigure}{0.49\linewidth}
	  \centering
	  \includegraphics[width=\linewidth, clip, trim={0em 1em 0em 1em}]{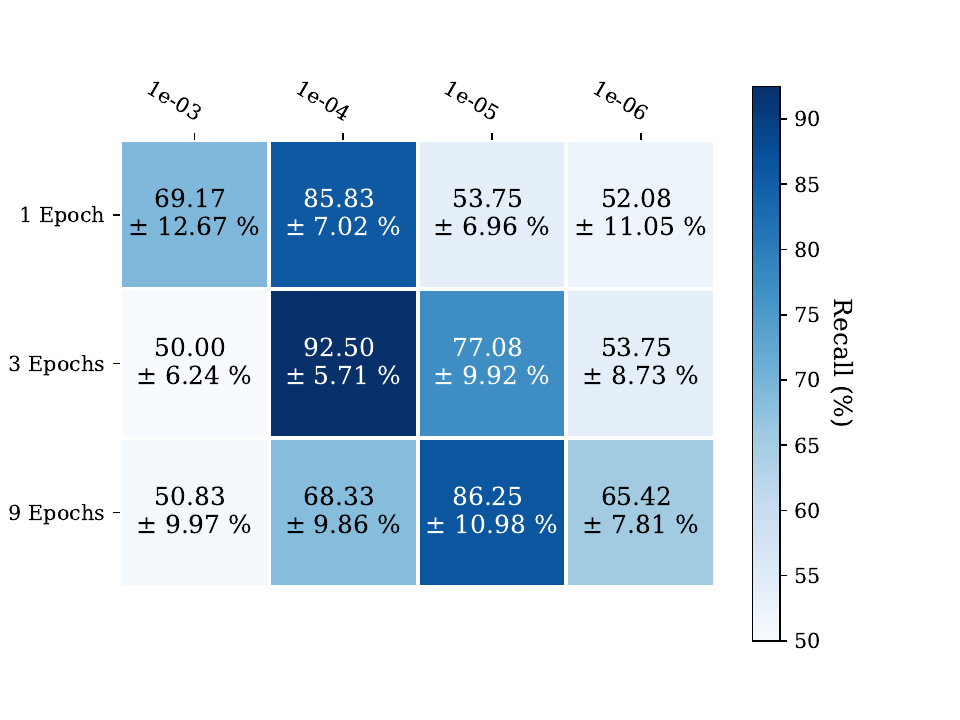}
	  \caption{Recall}
	\end{subfigure}
        \begin{subfigure}{0.49\linewidth}
	  \centering
	  \includegraphics[width=\linewidth, clip, trim={0em 1em 0em 1em}]{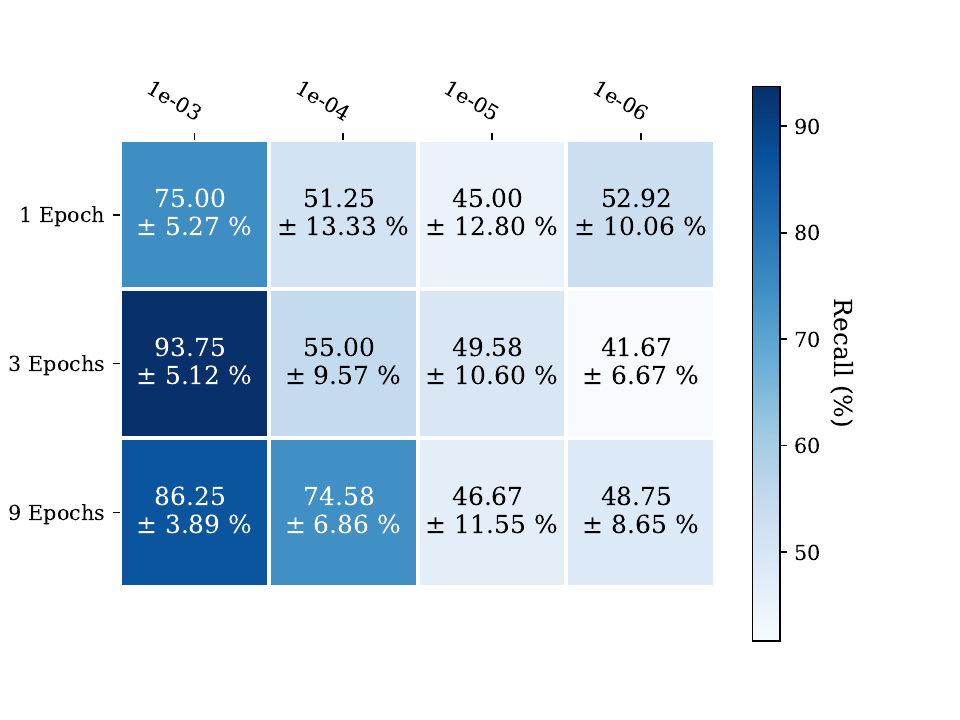}
	  \caption{Recall}
	\end{subfigure}
	\begin{subfigure}{0.49\linewidth}
	  \centering
	  \includegraphics[width=\linewidth, clip, trim={0em 1em 0em 0em}]{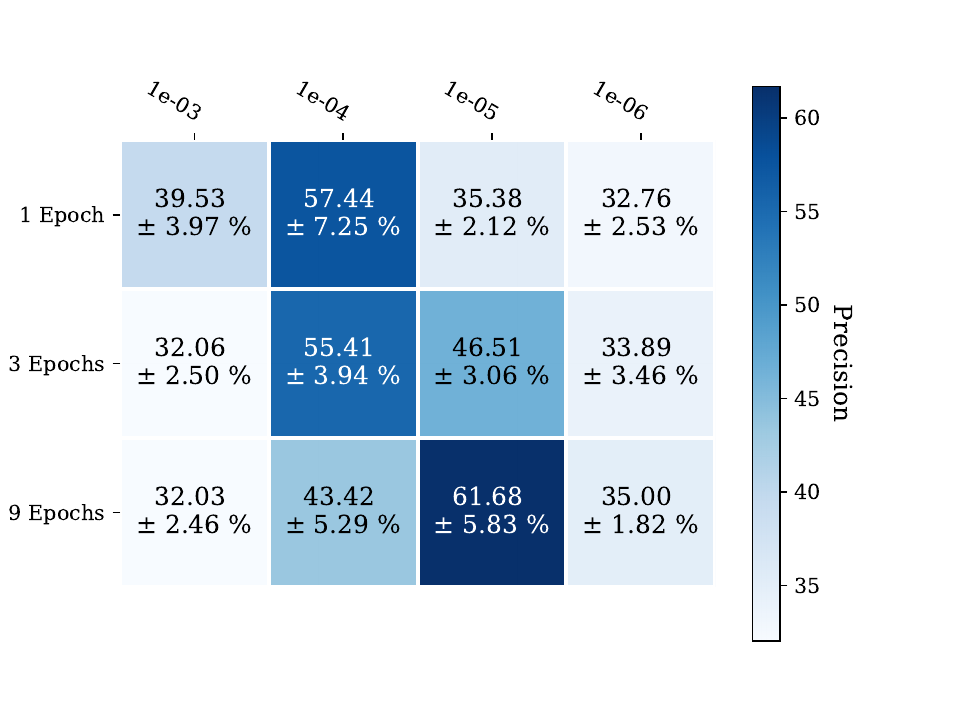}
	  \caption{Precision}
	\end{subfigure}
 	\begin{subfigure}{0.49\linewidth}
	  \centering
	  \includegraphics[width=\linewidth, clip, trim={0em 1em 0em 0em}]{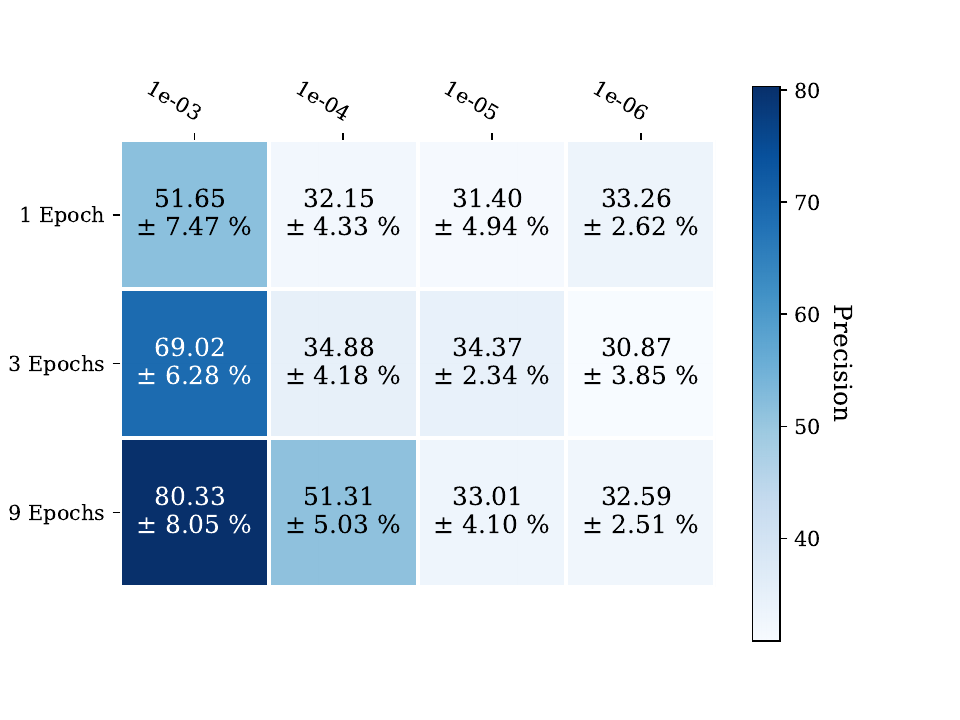}
	  \caption{Precision}
	\end{subfigure}
	\begin{subfigure}{0.49\linewidth}
	  \centering
	  \includegraphics[width=\linewidth, clip, trim={0em 1em 0em 0em}]{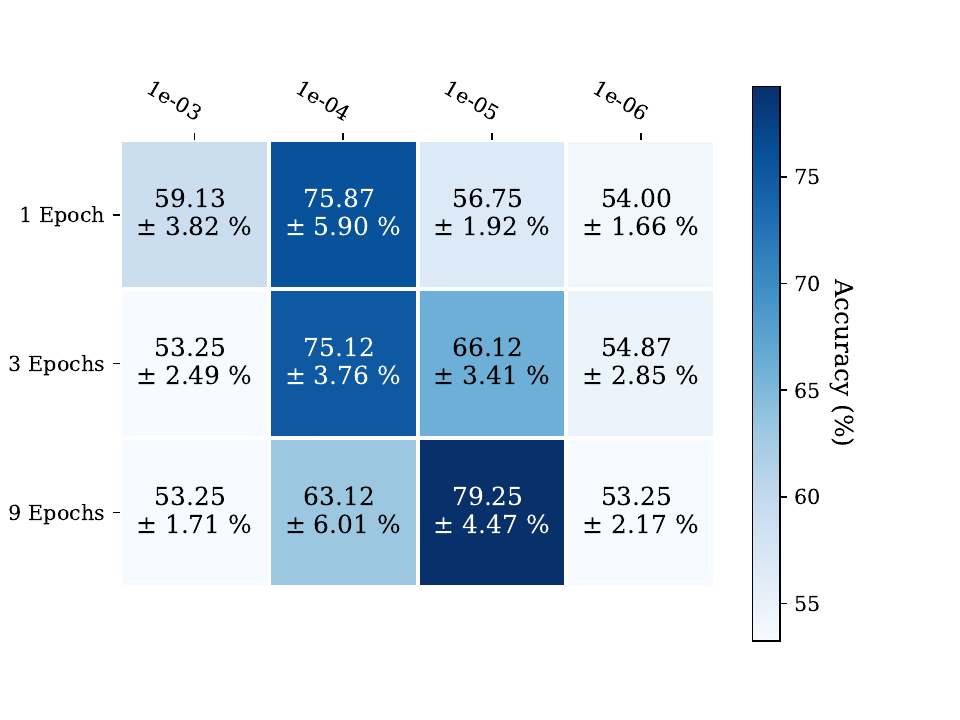}
	  \caption{Accuracy}
	\end{subfigure} 
	\begin{subfigure}{0.49\linewidth}
	  \centering
	  \includegraphics[width=\linewidth, clip, trim={0em 1em 0em 0em}]{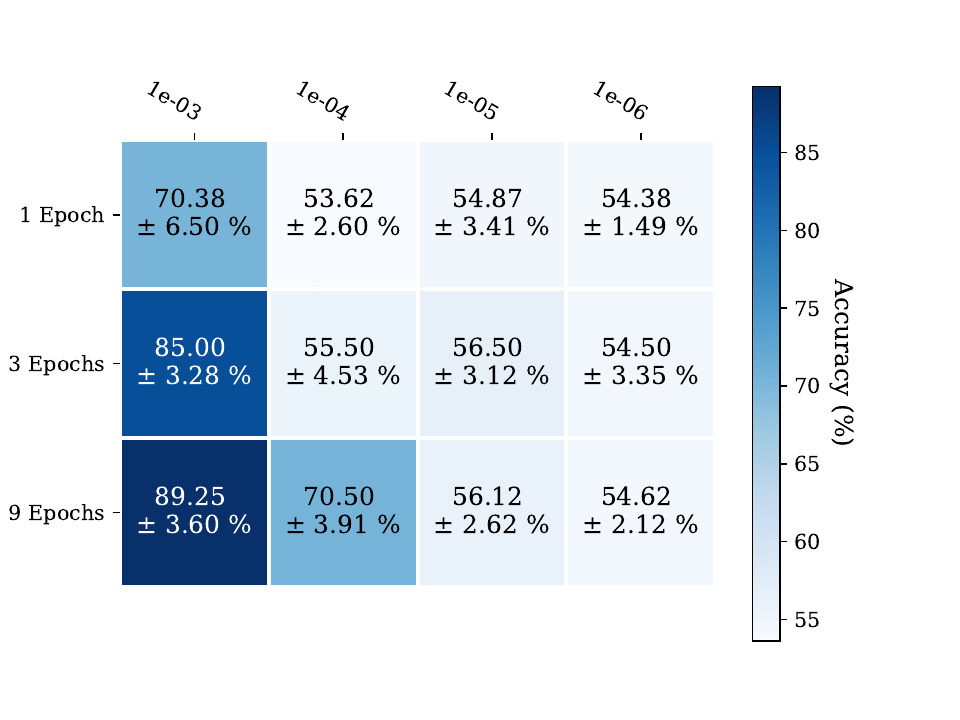}
	  \caption{Accuracy}
	\end{subfigure} 
	\caption{Recall (top), Precision (middle), and Accuracy(Bottom) on CIFAR 100 (left), and CIFAR 10 (right), non-IID, for different parameter pairs of learning rate and epoch count.}
\label{fig:non-iid c100}
\end{figure*}

We have conducted initial testing on CIFAR10 and CIFAR100 to explore the impact of various parameters on our model's performance. Our results, illustrated in Figures,~\ref{fig:iid c10}, and~\ref{fig:non-iid c100} demonstrate the effect of both learning rate and number of epochs on filtration performance.

We observe a balance between recall and accuracy that varies based on the parameters. This balance can be seen in both the CIFAR10 and CIFAR100 datasets. Additionally, the best parameters for IID and Non-IID may differ. For instance, the best recall for Non-IID and IID is achieved with different parameter pairs, and CIFAR100 also has a distinct parameter pair for IID compared to Non-IID.

\begin{figure*}[t]
	\centering
	\begin{subfigure}{0.49\linewidth}
	  \centering
	  \includegraphics[width=\linewidth, clip, trim={0em 1em 0em 1em}]{cifar10_test_2_recall_non_iid.pdf}
	  \caption{Recall}
	\end{subfigure}
	\begin{subfigure}{0.49\linewidth}
	  \centering
	  \includegraphics[width=\linewidth, clip, trim={0em 1em 0em 1em}]{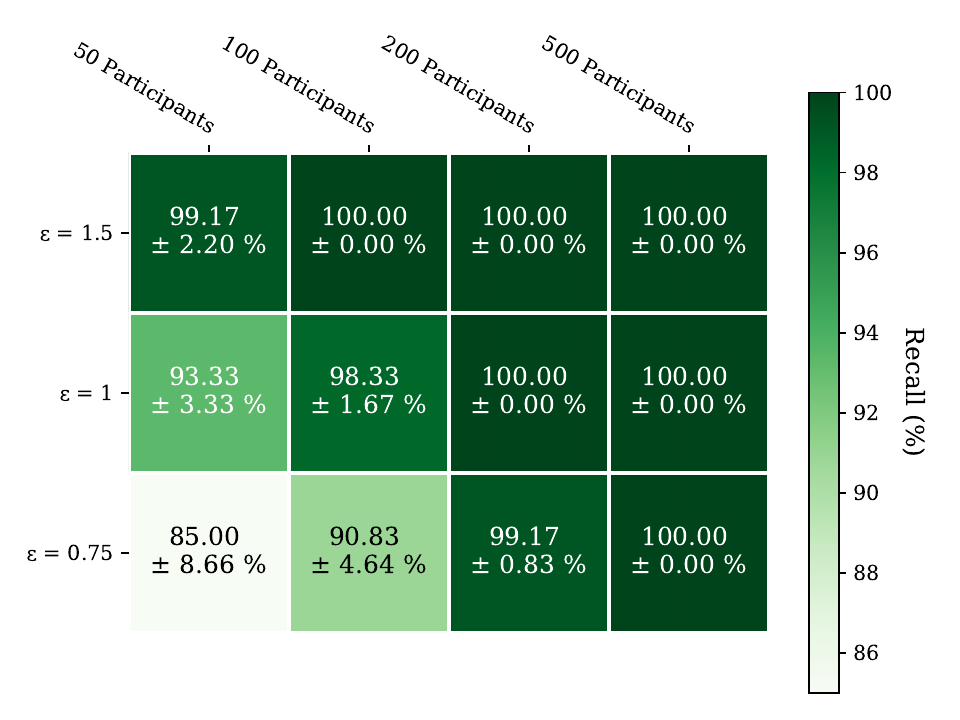}
	  \caption{Recall}
	\end{subfigure}
	\begin{subfigure}{0.49\linewidth}
	  \centering
	  \includegraphics[width=\linewidth, clip, trim={0em 1em 0em 0em}]{cifar10_test_2_precision_non_iid.pdf}
	  \caption{Precision}
	\end{subfigure}
 	\begin{subfigure}{0.49\linewidth}
	  \centering
	  \includegraphics[width=\linewidth, clip, trim={0em 1em 0em 0em}]{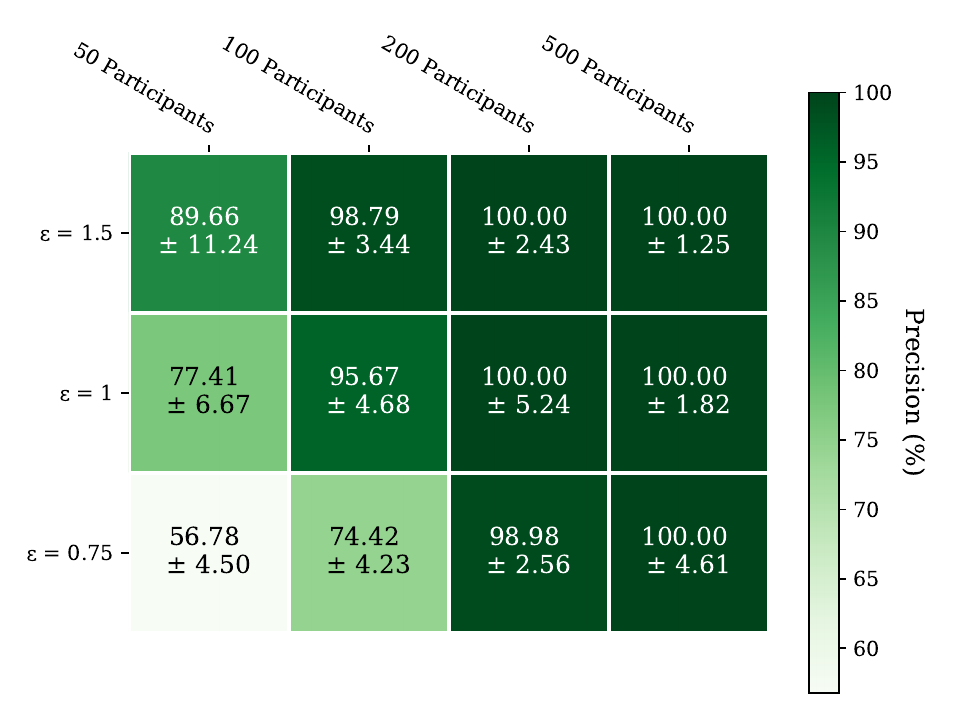}
	  \caption{Precision}
	\end{subfigure}
	\begin{subfigure}{0.49\linewidth}
	  \centering
	  \includegraphics[width=\linewidth, clip, trim={0em 1em 0em 0em}]{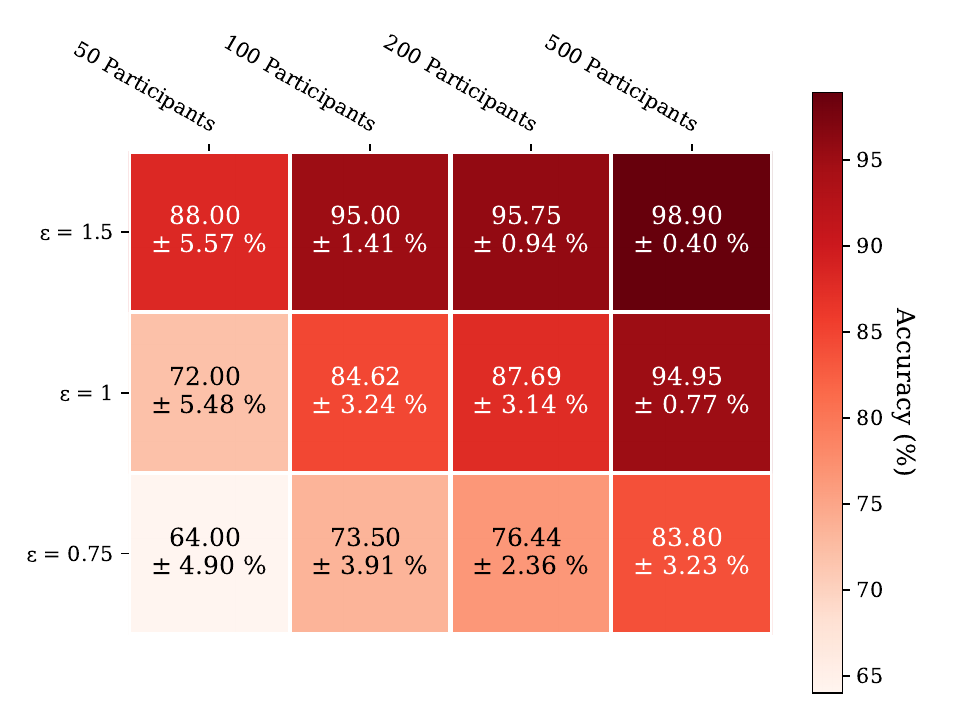}
	  \caption{Accuracy}
	\end{subfigure} 
 	\begin{subfigure}{0.49\linewidth}
	  \centering
	  \includegraphics[width=\linewidth, clip, trim={0em 1em 0em 0em}]{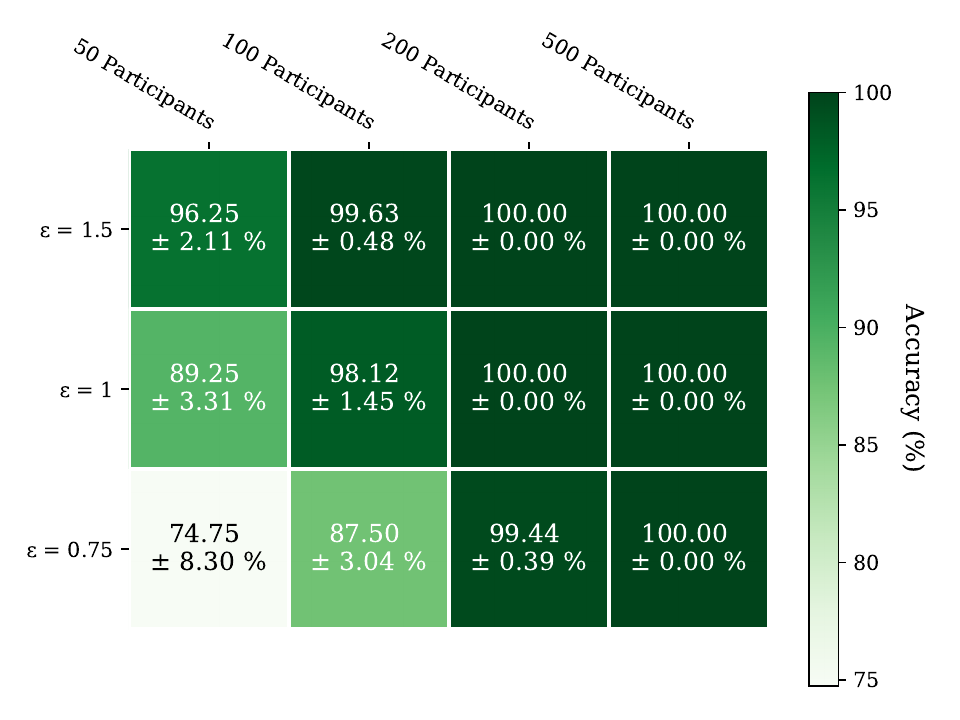}
	  \caption{Accuracy}
	\end{subfigure} 
	\caption{Recall (top), Precision (middle), and Accuracy(Bottom) on CIFAR 10, non-IID (left), IID (right), for increasing problem size (number of participants), and varying privacy guarantees ($\varepsilon$ -- lower $\varepsilon$ provides stronger privacy).}
\label{fig:non-iid epsilon full}
\end{figure*}

Finally, we examine the impact of various privacy guarantees ($\varepsilon$) and larger problem dimensions in Figure \ref{fig:non-iid epsilon full}. Our findings show that a smaller federation is needed to achieve the same level of performance when data is IID, compared to when it is Non-IID.

\end{document}